\documentclass[twocolumn]{aa}
\usepackage{graphicx}
\usepackage{natbib}
\def\vev#1{\langle#1\rangle}

\def\pmb#1{\setbox0=\hbox{#1}%
 \kern-.025em\copy0\kern-\wd0
 \kern.05em\copy0\kern-\wd0l
 \kern-.025em\raise.0433em\box0}
\def\eg{{\it e.g.\ }}

\def\ie{{\it i.e.\ }}

\def\del{\delta}                
\def\delg{\delta_g}           
\def\sig{\sigma}              
\def\bl{b_L}                      
\def\hpc{$h^{-1}$Mpc }
\def\hpcp{$ h^{-1}$Mpc. }
\def\hpcv{$ h^{-1}$Mpc, }

\def\delt{\delta_{T}}
\def\tdelt{{\tilde\delta}_{T}}

\def\tdelf{{\tilde\delta}_{F}}
\def\delo{\delta_{O}}
\def\tdelo{{\tilde \delta}_{O}}
\def\teps{{\tilde\epsilon}}
\def\fk{{\tilde F}_k}
\def\Wk{W_{k}}
\def\fskp{{\tilde F}^*_{k'}}
\def\Wskp{W^*_{k'}}

\def\z{$\,${\it z}$\,$}

\def\blx{{\bf x}}
\def\bly{{\bf y}}
\def\blr{{\bf r}}
\def\blk{{\bf k}}
\def\blkp{{\bf k'}}

\def\eg{{e.g.,~}}
\def\ie{{i.e.,~}}
\def\lsim{\raise0.3ex\hbox{$<$}\kern-0.75em{\lower0.65ex\hbox{$\sim$}}} 
\def\gsim{\raise0.3ex\hbox{$>$}\kern-0.75em{\lower0.65ex\hbox{$\sim$}}} 
\def\lesssim{\mathrel{\hbox{\rlap{\hbox{\lower4pt\hbox{$\sim$}}}\hbox{$<$}}}}
\def\gtrsim{\mathrel{\hbox{\rlap{\hbox{\lower4pt\hbox{$\sim$}}}\hbox{$>$}}}}
\def\sc1{\raise2.1ex\hbox{\tiny $r\!\!=\!\!4$}\kern-0.95em{\hbox{$=$}}}

\newcommand{\unit}[1]{\ifmmode \:\mbox{\rm #1}\else \mbox{#1}\fi}

\def\ltsima{$\; \buildrel < \over \sim \;$}
\def\simlt{\lower.5ex\hbox{\ltsima}}
\def\gtsima{$\; \buildrel > \over \sim \;$}
\def\simgt{\lower.5ex\hbox{\gtsima}}          

\def\sc{{\rm Science\ }}

\begin{document}

\title{The VIMOS VLT Deep Survey\thanks{Based on data
         obtained with the European Southern Observatory Very Large
         Telescope, Paranal, Chile }
}

\subtitle{Evolution of the non-linear galaxy bias up to z=1.5}

\author{
C. Marinoni \inst{1,2},
O. Le F\`evre \inst{2},
B. Meneux \inst{2},
A. Iovino \inst{1},
A. Pollo \inst{3},
O. Ilbert \inst{2,5},
G. Zamorani  \inst{5},
L. Guzzo \inst{3},
A. Mazure \inst{2},
R. Scaramella \inst{4},
A. Cappi \inst{5},
H.J. McCracken \inst{6},
D. Bottini \inst{7},
B. Garilli \inst{7},
V. Le Brun \inst{2},
D. Maccagni \inst{7},
J.P. Picat \inst{8},
M. Scodeggio \inst{7},
L. Tresse  \inst{2},
G. Vettolani \inst{9},
A. Zanichelli \inst{9},
C. Adami \inst{2},
S. Arnouts \inst{2}
S. Bardelli \inst{5},
J. Blaizot \inst{2},
M. Bolzonella \inst{10},
S. Charlot \inst{6,11},
P. Ciliegi \inst{9},
T. Contini \inst{8},
S. Foucaud \inst{7},
P. Franzetti \inst{7},
I. Gavignaud \inst{8},
B. Marano \inst{10},
G. Mathez \inst{8},
R. Merighi \inst{5},
S. Paltani \inst{2},
R. Pell\`o \inst{8},
L. Pozzetti \inst{5},
M. Radovich \inst{12},
E. Zucca  \inst{5},
M. Bondi \inst{9},
A. Bongiorno \inst{10},
G. Busarello \inst{12},
S. Colombi \inst{6},
O. Cucciati \inst{1,13},
F. Lamareille \inst{8},
Y. Mellier \inst{6},
P. Merluzzi \inst{12},
V. Ripepi \inst{12},
D. Rizzo \inst{8}
}

\offprints{C. Marinoni, e-mail: marinoni@brera.mi.astro.it}

\institute{
INAF - Osservatorio Astronomico di Brera, via Brera 28, 20121 Milano, Italia
\and
Laboratoire d'Astrophysique de Marseille, UMR 6110 CNRS-Universit\'e de Provence, Traverse du Siphon-Les trois Lucs, 13012 Marseille, France
\and
INAF - Osservatorio Astronomico di Brera, via Bianchi 46, 23807 Merate, Italia
\and
INAF - Osservatorio Astronomico di Roma, via Osservatorio 2, I-00040 Monteporzio Catone (Roma), Italia
\and
INAF - Osservatorio Astronomico di Bologna, via Ranzani 1, 40127 Bologna, Italia
\and
Institut d'Astrophysique de Paris, UMR 7095, 98 bis Bvd. Arago, 75014 Paris, France
\and
INAF - IASF,  Via Bassini 15, I-20133 Milano, Italia
\and
Laboratoire d'Astrophysique - Observatoire Midi-Pyr\'en\'ees, Toulouse, France
\and
INAF - Istituto di Radio-Astronomia, Via Gobetti 101, I-40129 Bologna, Italia
\and
Universit\`a di Bologna, Dipartimento di Astronomia, via Ranzani 1, 40127 Bologna, Italia
\and
Max Planck Institut fur Astrophysik, 85741 Garching, Germany
\and
INAF - Osservatorio Astronomico di Capodimonte, via Moiariello 16, 80131 Napoli, Italia
\and
Universit\`a di Milano-Bicocca, Dipartimento di Fisica, Piazza della scienza 3, 20126 Milano, Italia
}

\authorrunning{Marinoni et al.}
\titlerunning{The VVDS: Galaxy biasing up to z=1.5}

\date{June 14, 2005}

\abstract{

We present    the  first measurements  of   the  Probability 
Distribution Function (PDF) of galaxy fluctuations in the four-passes,
first-epoch   VIMOS-VLT    Deep    Survey   (VVDS) cone,  covering
0.4$\,$x$\,$0.4 deg between  0.4$<$\z$<$1.5.  We show that the PDF
of  density contrasts of the VVDS galaxies is an   unbiased  tracer of the underlying
parent distribution   up   to redshift  \z=1.5,   on  scales  R=8  and
10 \hpc. The second moment of the PDF,  \ie the {\it rms} fluctuations
of the galaxy density  field, is with   good approximation constant over  the
full redshift baseline investigated: we find  that, in redshift space,
$\sigma_8$ for galaxies  brighter than $\mathcal{M}_B^{c}=-20+5\log h$
has a   mean   value   of 0.94$\pm0.07$  in  the   redshift   interval
0.7$<$\z$<$1.5.  The third   moment, \ie the  skewness,
increases with   cosmic time: we find  that  the probability of having
underdense  regions   is  greater  at   \z$\sim$0.7   than it was   at
\z$\sim$1.5. By comparing the PDF of galaxy density contrasts with the
theoretically  predicted    PDF of mass   fluctuations    we infer the
redshift-, density-,  and  scale-dependence  of the biasing   function
$b(z, \del, R)$ between galaxy and matter overdensities up to redshift
\z=1.5.  Our  results can be summarized as follows:
i) the galaxy bias is an increasing function of redshift:
evolution is marginal up to \z$\sim$0.8 and more pronounced  
for \z$\gtrsim$0.8;
ii) the formation of bright galaxies
is inhibited  below a characteristic mass-overdensity threshold whose amplitude 
increases with redshift and luminosity;
iii) the biasing  function is  non linear
in  all the redshift bins investigated  with non-linear effects of the
order of a few to  $\sim 10\%$ on scales  $>5$ \hpcp By subdividing the
sample according  to galaxy luminosity and  colors,  we also show that: 
iv) brighter galaxies are   more strongly biased than  less  luminous
ones at every redshift and the dependence of  biasing on luminosity at
\z$\sim$0.8 is  in good  agreement with what  is observed  in  the local
Universe;  v) red objects are systematically more biased than blue
objects at all cosmic epochs investigated, but 
the relative bias between red and blue
objects is constant as a function  of  redshift in the interval  0.7$<$\z$<$1.5,
and its value ($b^{rel}\sim  1.4$) is similar to  what is  found at
\z$\sim$0.


}

\maketitle

\section{Introduction}
The understanding of how matter structures grow via gravitational
instability in an expanding Universe is quite well developed and has
led to a successful and predictive theoretical framework
\citep[\eg][]{pee80,dav85}.

One of the most critical problems, however, is to understand the
complex mechanisms which, on various cosmological scales, regulate the
formation and the evolution of luminous structures within the
underlying dark-matter distribution.  Its solution ultimately relies
on the comprehension of the ``microscopic'' physics which describes
how the baryons fall, heat-up, virialize, cool and form stars in the
potential wells generated by the dominant mass component of the
Universe, \ie the non-baryonic dark matter \citep[\eg][]{wr78}.  A
zero-order, minimal approach to this investigation consists in
characterizing ``macroscopically'' the cosmological matter
fluctuations in terms of a reduced set of fundamental quantities,
essentially their positions and mass scales, and in studying how the
respective spatial clustering and density amplitudes relate to the
corresponding statistics computed for light fluctuations. This
comparison scheme is generally referred to as matter-galaxy biasing
\citep[\eg][]{del99}.

An operational definition of bias  is conventionally given in terms of
continuous density  fields    by  assuming  that  the  local   density
fluctuation  pattern    traced   by galaxies  ($\delta_g$)    and mass
($\delta$)  are  deterministically  related via  the  ``linear biasing
scheme"
 
\begin{equation} \delg (\del)=b \del \label{lbs} \end{equation}

\noindent where the constant ``slope"  b is the biasing parameter \citep{kai84}.

This specific formulation, however, represents a very crude
approximation which is not based on any theoretical or physical
motivation. It is obvious, for example, that such a model cannot
satisfy the physical requirement $\delg(-1)=-1$ for any arbitrary
value $b\neq 1$.  In particular the biasing process could be non local
\citep[\eg][]{cat98}, stochastic \citep[\eg][]{del99} and non linear
\citep[\eg][]{mow96}. Moreover, both theory and numerical simulations
predict that the bias grows monotonically from the present cosmic
epoch to high redshifts \citep[\eg][]{dek87, fry96,mow96, teg98, bas01}.

From a theoretical perspective, light does not follow the matter 
distribution on
sub-galactic scales, where nearly 90$\%$ of 
dense, low-mass  dark matter 
fluctuations (M$\sim$ 10$^7$-10$^{8}$M$_{\odot}$)
failed to form stars and to become galaxies \citep[\eg][]{klyp99,moo99,dalk02}. 
A difference in the spatial distribution of visible and 
dark matter is predicted  also on galactic scales, since
the radial scaling of density profiles of dark
matter halos \citep{NFW} differs from the three-dimensional radial
distributions of light (Sersic and Freeman laws).  Galaxy biasing is
theoretically expected  also on cosmological scales.  
In particular, simulations of the large-scale structure 
predict the existence of a
difference in the relative distribution of mass and dark halos
\citep[\eg][]{cen92, bag98, kra99} or galaxies \citep[\eg][]{evr94,
bla00, kay01}. 
Various physical mechanisms for biasing have been
proposed, such as, for example, the peaks-biasing 
scheme \citep{kai84, bar86}, the probabilistic 
biasing approach \citep{col93}, or the biasing
models derived in the context of the extended Press \& Schechter
approximation \citep{mow96, mat98}.

Turning to the observational side, the fact that, in
the local Universe, galaxies cluster differently according to 
morphological type \citep{daghe76}, surface brightness \citep{davjo85},  
luminosity \citep{mau87}, or internal dynamics \citep{wtd88}
implies that not
all can simultaneously trace the underlying distribution of mass, and
that galaxy biasing not only exists, but  might also be sensitive 
to various physical processes.
As a matter of fact, redshift information which recently
became available for large samples of galaxies have significantly
contributed to better shaping our current understanding of galaxy
biasing, at least in the local Universe.  The analysis of the power
spectrum \citep{lah02} and bi-spectrum \citep{ver02} of the 2dF Galaxy
Redshift Survey \citep[2dFGRS][]{col01}
consistently shows that a flux-limited sample of local
galaxies (\z$<$0.25), optically selected in the $b_J$-band ($b_J \leq 19.4$),
traces the mass, \ie  it is unbiased, on  scales 5$<$R(\hpc)$<$30.

The galaxy correlation function has been measured up to redshift
$\sim$1 by the CFRS \citep{lef96},
and by the CNOC \citep{car00} surveys
giving conflicting evidence on clustering amplitude and bias
evolution \citep[see][]{sma}.  More recently, the analysis of
the first season DEEP2 data \citep{coil} seems to indicate that a
combined R-band plus color selected sample is unbiased at z$\sim$1.
On the contrary, measurements of the
clustering \citep{ste98, gia98, fou03} or of the amplitude of the
count-in-cell fluctuations \citep{ald98} of Lyman-break galaxies
(LBGs) at \z$\sim$3 suggest  that these objects are
more highly biased tracers of the mass density field than are galaxies
today.  Higher redshift domains have been probed by using photometric
redshift information \citep{arn99}, or compilation of heterogeneous
samples \citep{mag00}. Again, the clustering signal appears to come
from objects which are highly biased with respect to the underlying
distribution of mass.

While there is general observational consensus on the broad picture,
\ie that biasing must decrease with cosmic time, the elucidation of
the finer details of this evolution as well as any meaningful
comparison with specific theoretical predictions is still far from
being secured.  Since clustering depends on morphology, color and
luminosity, and since most  high redshift samples have been
selected according to different colors or luminosity criteria, it is
not clear, for example, how the very different classes of objects
(Ly-break galaxies, extremely red objects or ultraluminous galaxies),
which populate different redshift intervals, can be considered a
uniform set of mass tracers across different cosmic epochs.  
Furthermore, one must note that the biasing relation is likely
to be nontrivial, \ie non-linear and scale dependent, especially at
high redshift \citep[\eg][]{som01}.

Only large redshift surveys defined in terms of uniform selection
criteria and sampling typical galaxies (or their progenitors),
rather than small subclasses of peculiar objects, promise
to yield a more coherent picture of biasing evolution.  In
particular, the 3D spatial information provided by the VIMOS-VLT Deep
Survey \citep[VVDS,][hereafter Paper I]{lef04a}  should allow us to investigate the mass and scale
dependence, as well as to explore the time evolution of the biasing
relation between dark matter and galaxies for a homogeneous,
flux-limited (I$\leq$24) sample of optically selected galaxies.

The intent of this paper is to provide a measure, on some
characteristic scales R, over the continuous redshift interval  0.4$<$\z$<$1.5, 
of the {\it local}, {\it non-linear}, {\it deterministic} biasing function

\begin{equation} b=b(z, \del, R) \label{nlbp}. \end{equation}

The goal is to provide an observational benchmark to theories
predicting the efficiency of structure formation, or 
semi-analytical simulations of galaxy evolution.  The problem,
however, is to find an optimal strategy to evaluate the biasing
function in the quasi-pencil-beam geometry of the first-epoch VVDS survey.
At present, the angular size of the first-epoch VVDS
redshift cone ($\sim 0.5$deg$^2$) does not allow  
us to constrain the biasing function
using high order moments of the galaxy distribution (for example the 3-point 
correlation function). 
Moreover, we
cannot determine the biasing function simply regressing the galaxy
fluctuations ($\delg$) versus mass fluctuations ($\del$).  The
fundamental limitation preventing such an intuitive comparison is
evident: it is easy to ``pixelize" the survey volume and to measure
the galaxy fluctuations in each survey cell, but, since the VVDS is not
a matter survey, it is much less straightforward to assign to each cell a
mass density value.
Thus, in this paper, we  take an
orthogonal approach and we infer the  biasing relation $\delg=\delg(\del)$ between 
mass and galaxy overdensities from their respective probability distribution
functions (PDFs) $f(\del)$ and $g(\delg)$:
assuming a one-to-one mapping between mass and galaxy overdensity 
fields, conservation of probability implies

\begin{equation}
\frac{d\delg(\del)}{d\del}=\frac{f(\del)}{g(\delg)}.\label{probc}
\end{equation}

The advantage over other methods is that we can
explore the functional form of the relationship $\delg=b(z,\del, R)
\del$ over a wide range in mass density contrasts, redshift intervals
and smoothing scales R without specifying any {\it a-priori}
parametric functional form for the biasing function.  

In pursuing our approach, we assume that 
the PDF of matter overdensities
$f(\del)$ is satisfactorily described by theory and N-body simulations.
What we will try to
assess explicitly, is the degree at which  the measured PDF of the
VVDS overdensities $g(\delg)$ reproduces the PDF of the underlying parent
population of galaxies.  The large size and high redshift sampling
rate of the VVDS spectroscopic sample, together with the multi-color
information in the B, V, R, I filters of the parent photometric
catalog and the relatively simple selection functions of the survey,
allow us to check for the presence of observational systematics
in the data. In principle, this analysis helps us to constrain
the range of the parameter space where first-epoch VVDS data can be
analyzed in a statistically unbiased way and results can be  meaningfully
interpreted.

The outline of the paper is the following: in \S 2 we briefly describe the
first-epoch VVDS data sample. In \S 3 we introduce the technique
applied for reconstructing the three-dimensional  density field traced 
by VVDS galaxies, providing details about corrections for various 
selection effects. In \S 4 we  outline the construction of the
PDF of galaxy overdensities and test its statistical
representativity. We then derive the PDF of VVDS density contrasts
and analyze its statistical moments. 
In \S 5 we review the theoretical properties of the
analogous statistics for mass fluctuations. 
Particular emphasis is
given to the problem of projecting the mass PDF derived in real space
into redshift-perturbed comoving coordinates in the high redshift Universe.  
The method for
computing the biasing function is introduced and tested against possible
systematics in \S 6. VVDS results are presented and  discussed in \S 7.
and compared to 
theoretical models of biasing evolution in \S 8. 
Conclusions are drawn in \S 9.

The coherent cosmological picture emerging from independent 
observations and analysis motivate us to frame all the results presented 
in this paper in the context of a $\Lambda$CDM cosmological model
with $\Omega_m=0.3$ and $\Omega_{\Lambda}=0.7$.
Throughout, the Hubble constant is parameterized via  $h=H_{0}/100$.
All
magnitudes in this paper are in the AB system (Oke \& Gunn 1983), 
and from now on we will drop the suffix AB.

\section{The First-Epoch VVDS  Redshift Sample}

The primary observational goal of the VIMOS-VLT Redshift Survey 
as well as the survey strategy and first-epoch observations
in the VVDS-0226-04 field (from now on simply 
VVDS-02h) are presented in Paper I.

Here it is enough to stress that, 
in order to minimize selection biases, the VVDS survey in the VVDS-02h field, has
been conceived as a purely flux-limited ($17.5\leq I \leq24$) survey, \ie  no target
pre-selection according to colors or compactness is implemented.
Stars and QSOs have been {\it a-posteriori} removed from the final
redshift sample. Photometric data in this field
 are complete and free from surface brightness selection 
effects, up to the limiting magnitude I=24 \citep{mcc03}.

First-epoch spectroscopic observations in the VVDS-02h field 
were carried out using the VIMOS multi-object
spectrograph \citep{lef03} during two runs between October and 
December 2002 (see Paper I). 
VIMOS observations have been performed using 1 arcsecond wide
slits and the LRRed grism which covers the spectral range
$5500<\lambda(\AA)<9400$ with an effective spectral
resolution $R\sim 227$ at $\lambda=7500\AA$.  The accuracy in redshift measurements is
$\sim$275 km/s.  Details on 
observations and data reduction are given in Paper I, and in \citet{lef04}.

The first-epoch VVDS-02h data sample extends over a
sky  area of 0.7$\times$0.7 deg (which was targeted according
to a 1, 2 or 4 passes strategy, \ie giving to 
any single  galaxy in the field 1, 2 or 4 chances to be targeted by VIMOS masks
(see fig. 12 of paper I)) 
and has a median depth of about \z$\sim$0.76.
It contains 6582 galaxies with secure redshifts (\ie redshift determined 
with a quality flag$\ge$2 (see Paper I)) and probes a comoving volume (up to \z=1.5)
of nearly $1.5\cdot 10^6 h^{-3}$Mpc$^{3}$
in a standard $\Lambda$CDM cosmology. This volume 
has transversal dimensions $\sim$ 37x37 \hpc at \z=1.5 and extends 
over 3060 \hpc in radial direction.

For this study we define a sub-sample (VVDS-02h-4) 
with galaxies having redshift  \z$<$1.5 and selected in a continuous sky region
of 0.4$\times$0.4 deg which has been
homogeneously targeted four times by VIMOS slitmasks. 
Even if we measure
redshifts up to \z$\sim$5 and in a wider area, 
the conservative angular and redshift limits 
bracket the range where we can sample in a denser way
the underlying galaxy distribution and, thus,
minimize biases in the reconstruction of the density field (see the analysis in \S 4.1).
The VVDS-02h-4  subsample contains 3448 galaxies with secure redshift 
(3001 with $0.4<z<1.5$) and probes one-third of the total VVDS-02h volume.
This is the main sample used in this study.

\section{The Density Field Reconstruction Scheme}

The first ingredient we need in order to derive the biasing relation

\begin{equation} \delg=b(z,\del, R) \del \end{equation}

\noindent is a transformation scheme for diluting an intrinsic point-like
process, such as the  galaxy distribution in a redshift survey, 
into a continuous 3D overdensity field \citep[see the review by][]{sw}.  
We write the dimensionless density contrast at the {\it comoving} position {\bf r},
smoothed over a typical dimension $R$ as

\begin{equation} \displaystyle \delg({\bf r},R) =
\frac{\rho_{g}({\bf r},R)-\overline{\rho_g}}{\overline{\rho_g}}.
\label{defdg} \end{equation}

\noindent and we  define \citep[\eg][]{hud} the smoothed number
density of galaxies above the absolute magnitude threshold $\mathcal{M}^c$ as
the convolution between Dirac's delta functions and some arbitrary
filter

\begin{equation} \displaystyle \rho_{g}({\bf
r},R,<\mathcal{M}^c)=\sum_i \frac{ \del^{D}({\bf r}-{\bf
r_i})*F\big(\frac{|{\bf r-r_i}|}{R}\big)}{S(r_i, \mathcal{M}^c)\Phi_z(m)}
\label{defrg} \end{equation} 

\noindent where the sum is taken over all the
galaxies in the sample, $S(r, \mathcal{M}^c)$ is the distance-dependent
selection (or incompleteness) function of the sample (see \S 3.2),
$\Phi_z(m)$ is the redshift sampling function (see \S 3.3) and 
$F(|{\bf r}|/R)$ is a smoothing kernel of width $R$. In this paper, the
smoothing window F is
modeled in terms of a normalized Top-Hat (TH) filter

\begin{equation} \displaystyle F\Big(\frac{|{\bf r}|}{R}\Big)= \frac{3}{(4 \pi
R^3)} \Theta\Big(1-\frac{|{\bf r}|}{R}\Big),  \label {wt}
\end{equation}

\noindent where $\Theta$ is the Heaviside  function, defined as $\Theta(x) = 1$ for 
0 $\leq$ x $\leq$ 1, and $\Theta(x) = 0$ elsewhere.

Within this weighting scheme, shot-noise errors are evaluated  by
computing the variance of the galaxy field

\begin{equation} \displaystyle
\epsilon(\blr)=\frac{1}{\overline{\rho_g}}\Bigg[\sum_i
\Bigg(\frac{F\big(\frac{|{\bf r-r_i}|}{R}\big)}{S(r_i,\mathcal{M}^c)
\Phi_z(m_i)}\Bigg)^2\Bigg] ^{1/2} \label{shot} \end{equation}

Note that all coordinates are comoving and that  the mean density
$\overline{\rho_g}$ depends on  cosmic time.
Since we observe an evolution of nearly a  factor of two in the mean density of galaxies
brighter than $\mathcal{M}^*_{B}(z=0)$ from redshift z=0 up to z=1 (see Ilbert et al. (2005), hereafter Paper II), we compute the value of $\overline{\rho_g}$
at position r (corresponding to some look-back time t) with eq. (\ref{defrg}) by
simply averaging the galaxy distribution in survey slices $r\pm R_s$ where $R_s=400$ \hpcp
We note that conclusions drawn in this paper depend very weakly on the choice
of $R_S$ in the interval 200$<R_s$( \hpc)$<$600.

In this paper, the density field is evaluated at positions {\bf r} in
the VVDS-02h volume that can 
be either random or regularly displaced on a 3D grid (see section 3.4).  
Even if we correct for the  different sampling rate in the VVDS-02h field
(by weighting each galaxy by the inverse of redshift sampling function ($\Phi_z(m)$))
we always
select, for the purposes of our analysis,  only 
the density fluctuations  recovered in spheres having at least 70$\%$ of their volume 
in the denser 4-passes volume.  
This in order to minimize the Poissonian noise 
due to the sparser redshift sampling outside the VVDS-02h-4 field.

We also consider only volumes above the redshift threshold \z$_t$ where 
the transversal dimension L of the first-epoch VVDS-02h cone is L$(z_t)>$2R. 
As an example, for TH windows of size R=5(10) \hpc we have \z$_t\sim$ 0.4(0.7).
Within the redshift range 0.4$<$\z$<$1.5 the VVDS-02h field contains 5252 galaxies 
(of which 3001 are in the four passes region.)

Note that we have characterized the galaxy-fluctuation field in terms
of the number density contrast, instead of the luminosity density
contrast, because the former quantity is expected to show a time-dependent
variation which is more sensitive to the galaxy evolution history
(formation and merger rates, for example).  Moreover, as described in
section \S 3.4, a robust description of the density field and a
reliable determination of the PDF shape can be obtained only
minimizing the shot noise component of the scatter; this is more
easily done by considering galaxy number densities rather than galaxy
light densities.

The most critical elements of the smoothing process are directly
readable in eq. (\ref{defrg}): we must first
evaluate galaxy absolute magnitudes at each redshift 
in the most reliable way, then
specify the selection function and the redshift sampling rate of the
VVDS survey. In the next sections we will describe how these quantities 
have been evaluated.

\subsection{The K-correction}

The absolute magnitude is defined as:

\begin{equation} M^{r}=m^{o}-5\log d_{L}(z,\vec{\Omega})-K(z,{\rm
SED}) \label{Magabs} \end{equation}

\noindent where the suffixes $r$ and $o$ designate respectively the
rest-frame band in which the absolute magnitude is computed and the
band where the apparent magnitude is observationally measured, and
$d_L$ is the luminosity distance evaluated in a given {\it
a-priori} cosmology (\ie  using an appropriate set of cosmological
parameters $\vec{\Omega} \equiv [\Omega_m, \Omega_{\Lambda}])$.

The correction factor $K$, which depends on redshift and the spectral
energy distribution (SED), accounts for the fact that the system
response in the observed frame corresponds to a narrower, bluer
rest-frame passband, depending on the redshift of the observed object.
A complete description of the application of this transformation
technique to VVDS galaxies is detailed in Paper II. 

The estimate of the galaxy absolute luminosity is thus affected by  
the uncertainties introduced by probing redshift regimes
where the $k-correction$ term cannot be neglected.
Anyway,  using mock catalogs simulating the VVDS survey, we have shown   
(see Fig. A.1 in Paper II) that the errors  in the recovered 
absolute magnitude are significantly smaller in the  B band 
($\sigma_B=0.08$) than in the I band ($\sigma_I=0.17$).
Thus, in what follows we will use absolute
luminosities  determined in the B-band  rest-frame.

\subsection{The Radial Selection Function}

Since our sample is limited at bright and faint apparent magnitudes (17.5$\leq$I$\leq$24),  
at any given redshift we can only observe galaxies in a specific, redshift-dependent,  
absolute magnitude range. 
It is usual to describe the sample radial incompleteness by defining
the selection function in terms of the galaxy luminosity function
$\varphi(\mathcal{M})$

\begin{equation}
S(r,\mathcal{M}^{c})=\frac{\int_{\mathcal{M}_{b}(r)}^{\mathcal{M}_{f}(r)}
\varphi(\mathcal{M})d\mathcal{M}}{\int_{-\infty}^{\mathcal{M}^{c}}
\varphi(\mathcal{M})d\mathcal{M}}. \label{sul} \end{equation}

Here $\mathcal{M}_{b}(r)$ and $\mathcal{M}_{f}(r)$ are the B-band
absolute magnitudes which correspond, at distance r, to the I-band
limiting apparent magnitudes $m_{b}=17.5$ and $m_{f}=24$ respectively (see discussion
at the end of the section).  Since
eq. (\ref{Magabs}) also depends  on galaxy colors, we compute its mean value
at distance r by a weighted average over  the population mix 
observed at distance r.

The VVDS luminosity function (LF) has been derived in Paper II 
and is characterized
by a substantial degree of evolution over the redshift range
0$<$\z$<$1.5. Therefore, we  estimate 
$\varphi(\mathcal{M})$ in the B band at any given position in the redshift
interval [0, 1.5] by interpolating, with a low order
polynomial function, the Schechter shape parameters $\alpha$ and $\mathcal{M}^*$
given in  Table 1 of  Paper II. 

Assuming $\mathcal{M}_{B}^c=-15$ in eq. \ref{sul}, which corresponds
to the limiting absolute magnitude over which the LF of the VVDS-02h
sample can be robustly constrained in the lowest redshift bins, the
selection function exponentially falls by nearly 2 orders of magnitude
in the redshift range up to \z$<$1.5.  Thus, the density field
reconstruction strongly depends on the radial selection function used
especially at high redshifts, where eq. \ref{sul} can be affected by
possible systematics in the determination of the LF or in the
measurements of faint magnitudes.  Therefore, we will also analyze
volume-limited sub-samples, which are essentially free from these
systematics.

Since in a magnitude limited survey progressively brighter galaxies
are selected as a function of redshift, a volume-limited sample also
allows us to disentangle spurious luminosity-dependent effects from
the measurement of the redshift evolution of the biasing function.

\subsection{The Redshift Sampling Rate}

As for most redshift surveys, the VVDS does not target
spectroscopically all the galaxies which satisfy the given flux limit
criteria in the selected field of view (see Paper I). 
Because of the sparse sampling strategy, 
we have to correct the density estimator with a sampling rate weight $\Phi_z$ in
order to reconstruct the real
underlying galaxy density field in a statistically unbiased way.

\begin{figure} \includegraphics[width=8.7cm]{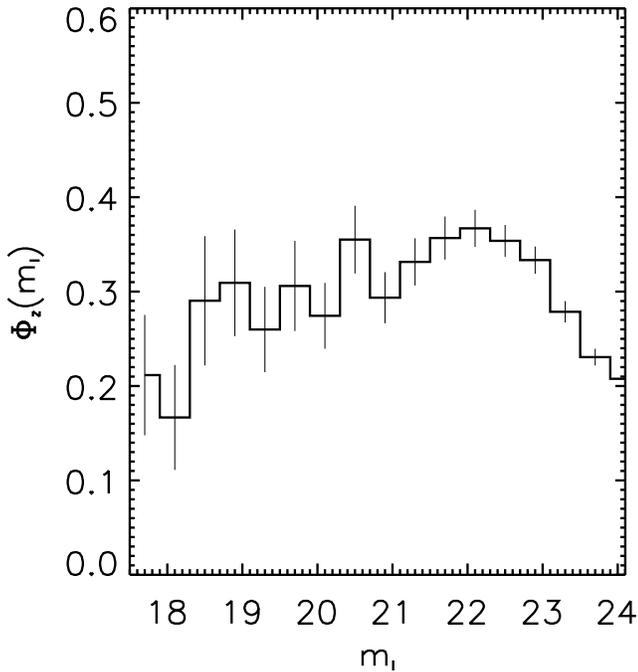}
\caption{The VVDS redshift sampling rate in the four-passes VVDS-02h-4 region 
is plotted versus the observed apparent magnitude in the
I-band. The mean redshift sampling rate is $\sim 0.3$.} 
\label{figsr} \end{figure}

The VVDS redshift sampling rate is the combination of two effects: i) only a
fraction of the galaxies ($\sim 40\%$, see Paper I) in the photometric sample is targeted ({\it target sampling rate}), ii) and only a fraction of the targeted
objects ($\sim 80\%$ see Paper I) yield a redshift ({\it spectroscopic success rate}).
We can
model this correction term, by assuming, in first approximation, that
the sampling rate depends only on the apparent magnitude. 
Since the VVDS targeting strategy is
optimized to maximize the number of slits on the sky, the selection of
faint objects is systematically favored.
Inversely, the ability of measuring a redshift degrades progressively
towards fainter magnitudes, \ie for spectra having lower
signal-to-noise ratios (the spectroscopic success rate 
decreases from $>90\%$ at I$\sim$22 down to $\sim 60\%$ at I$\sim$24.) 
These two opposite effects conspire 
to produce the magnitude-dependent sampling rate function shown in 
Fig. \ref{figsr}.
Clearly, with such an approximation, we neglect any possible dependence 
of the sampling rate from other important parameters such as for example 
surface brightness, spectral type or redshift.
However, in in \S 3.2 of Paper II  we showed, using photometric redshifts, that any 
systematic sampling bias introduced by a possible 
redshift dependence of the spectroscopic success rate is expected to 
affect only the tails of our observed redshift distribution (\z$<$0.5 and \z$>$1.5) \ie 
redshift intervals not considered in this study (see \S 3).

We describe the VVDS sampling completeness $\Phi_z(m)$ at a given
magnitude $m$, as the fraction of objects with measured redshifts
$N_{z}$ over the number $N$ of objects detected in the photometric
catalog.

\begin{equation} \Phi_z(m)=\frac{\sum_{i=1}^{N_{z}}w(m-m_i,
dm)}{\sum_{i=1}^{N}w(m-m_i, dm)} \end{equation} 

\noindent with the
window function $w$ defined as: 

\begin{equation} \label{eqn:W}
w(m-m_i, dm) = \left\{ \begin{array}{rl} 1 & \mbox{if $|m-m_i| <
dm/2 $} \\ 0 & \mbox{otherwise} \end{array} \right.  \end{equation}

Working in low resolution mode (i.e allowing up to 4 galaxies 
to have spectra aligned along the same
dispersion direction with  a typical sky separation of 2 arcmin) 
and observing the same region of sky 4
times for a total of about 16 hours (four-passes strategy), we
can measure redshifts for  nearly 30$\%$ of the parent photometric
population, as shown in Fig. \ref{figsr}.  
In other terms, on average, nearly one over three galaxies with
magnitude I$\leq$24 has a measured redshift in the four-passes VVDS
region.
This high spatial sampling rate is a critical factor
for minimizing biases in the reconstruction of the 3D density field of
galaxies.  In particular we note that 
our I$\leq$24, \z$<$1.5 VVDS-02h-4 sample is characterized by an  
effective  mean inter-particle separation in the redshift range [0,1.5]
($\vev{r} \sim 5.1$ \hpc) which is
smaller than that of the original CFA sample ($\vev{r}\sim 5.5$$ h^{-1}$Mpc)
used by \citet{dav81} to reconstruct the  3D density field 
of the local Universe (\ie out to $\sim$ 80  \hpc). 
At the median depth of the survey, \ie in the redshift interval 
0.75$\pm0.05$, the mean inter-particle separation is 4.4 \hpcv a value nearly equal  to the mean inter-particle separation at the median depth of the 2dFGRS.
We finally note that in the redshift range [0.7,1.35], which is also covered by
the DEEP2 survey, the VVDS mean-inter-particle separation is 5.5 \hpc
compared to the value $\sim$ 6.5 \hpc inferred from the values quoted by
Coil et al. (2004) for their most complete field, currently covering 0.32 deg$^2$.

By dividing the VVDS-02h-4 field in smaller cells and repeating the analysis,   
we conclude that the sampling rate does not show appreciable variations, \ie  
the angular selection  function can be considered constant for the purposes of 
our analysis. This corresponds to the fact that the success rate in redshift measurement
in each VIMOS quadrant  (\ie the {\it spectroscopic  success rate} per mask)
is, in good  approximation, constant and equal to $\sim$80$\%$ (see Paper I).

\subsection{Shot Noise}

\begin{figure}\centering \includegraphics[width=8.7cm, angle=0]{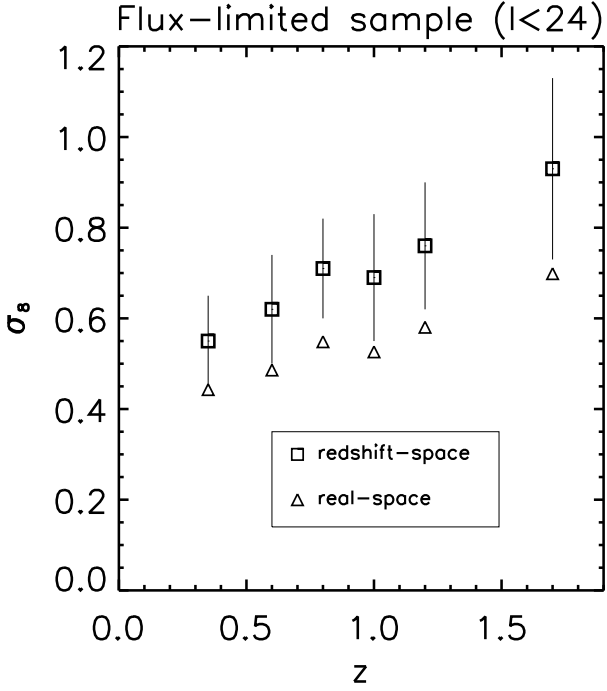}
\caption{The real- and redshift-space {\it rms} fluctuations of the flux-limited
VVDS sample recovered using eq. (\ref{pows})
and the results of the correlation function analysis presented in Paper III are
plotted  at six  different redshifts in the interval 0.4$<$\z$<$1.7.}
 \label{powsfig} \end{figure}

In a flux-limited sample, the shot noise in the density field is an
increasing function of distance (see eq. (\ref{shot})).
One may correct for the increase of the mean VVDS inter-particle
separation as a function of redshift (and thus the increase of the 
variance of the density field) by opportunely increasing the length 
of the smoothing window \citep[\eg][]{sw}. 
However, since we are interested in comparing the fluctuations recovered 
on the same scale at different redshifts in a flux limited survey,
we take into account the decreased sampling sensitivity of 
the survey at high redshift in an alternative way.

We 
deconvolve the signature of this noise from the density maps by
applying the Wiener filter technique \citep[cf.][]{pre92} which
provides the {\it minimum variance} reconstruction of the smoothed
density field, given the map of the noise and the {\it a priori}
knowledge of the underlying power spectrum \citep[\eg][]{lah94}.
The application of the Wiener denoising procedure to the specific
geometry of the VVDS sample is described in detail in appendix A.

Here we note that the Wiener filter requires a model for the underlying 3D power spectrum
P(k,z) which we compute, over the frequency scales where the correlation function 
of VVDS galaxies is well constrained ($0.06 \leq k \leq 10$),
 as (see eq. (\ref{pow}) in Appendix A):

\begin{equation} P(k,z)=4 \pi
\frac{{r_0(z)}^{\gamma(z)}}{k^{3-\gamma(z)}}\Gamma(2-\gamma(z))
\sin\frac{\gamma(z) \pi}{2}, \label{pows}\end{equation}

\noindent where the normalization $r_0(z)$ and the slope $\gamma(z)$
of the correlation function at redshift z have been derived by
interpolating the values measured in various redshift intervals of
the VVDS-02h volume by \citet{lef05b}(hereafter Paper III).

\begin{figure*} \centering \includegraphics[width=18.0cm]{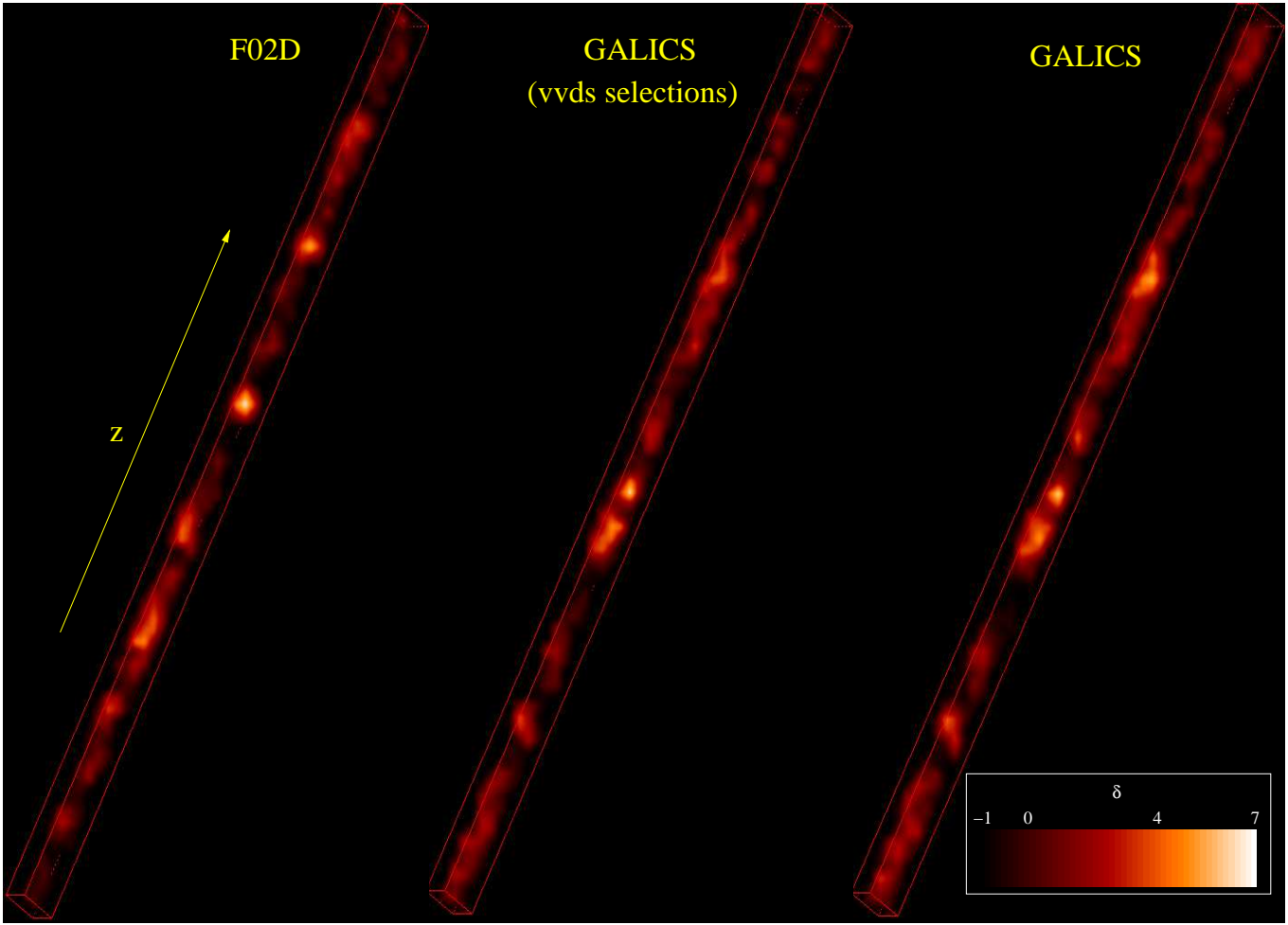}
\caption{ 3D density field traced by the galaxy distribution in the
VVDS-02h Field ({\em left}, 1641 galaxies), in the flux-limited (I$\leq$24) GALICS simulation ({\em
right}, 9450 galaxies), and in the flux-limited GALICS sample  after applying the VVDS
target selection criteria ({\em center}, 1656 galaxies). Data span the redshift interval [0.8,1.1].  
In each cone,  the galaxy distribution
is continuously smoothed using a TH window function with R=5 \hpc
which nearly corresponds to the mean inter-particle
separation in this same redshift interval.
The cone metric was computed assuming a $\Lambda$CDM 
cosmology and the correct axis ratio between transversal and radial
dimensions has been preserved.  The cones have approximate transverse
dimensions of 28 \hpc at \z=1 and extend over 527 \hpcp}  \label{figco}
\end{figure*}

The variance of the 
galaxy distribution on a 8 \hpc scale  in the VVDS-02h sample can be obtained 
by integrating eq. (\ref{pows})
using a TH window of radius 8 \hpc
and the ($r_0,\gamma$) parameters of the VVDS correlation function

\begin{equation}
\sigma_8^2=\frac{1}{2\pi^2}\int P(k)\fk^2 k^2dk \label{vari}
\end{equation}

\noindent where $\fk$ is the Fourier transform of the TH filter 
(see eq. (\ref{ftth}) in Appendix A).
Note that, by integrating the power spectrum 
down to $k \rightarrow 0$, 
\ie extrapolating the power law shape of eq.
(\ref{pows}) beyond $L \sim 100$ \hpc ($k\lesssim$0.06), one would revise upwards 
the value of $\sigma_8$ by $\sim 2\%$(at z=0.35) and by $\sim 4\%$ at z=1.4. Since, however, 
the amplitude of the power spectrum on large scales is expected 
to downturn and to be systematically lower than predicted by eq. (\ref{pows}), we safely 
conclude that,  with
our computation scheme, the inferred $\sigma_8$ value should be biased low
by no more than $\sim 2\%$ and $4\%$ in the first and last redshift bin, respectively. 

Projecting the results from real-space into the redshift-distorted  space (see \S 5),
\ie implementing the effects of large-scale streaming motions, 
we infer that the {\it rms} of galaxy fluctuations are 
$\sigma_8$=[0.55$\pm$0.10, 0.62$\pm$0.12, 0.71$\pm$0.11, 0.69$\pm$0.14, 0.75$\pm$0.14, 0.92$\pm$0.20] at redshift z=[0.35,0.6,0.8,1.2,1.2,1.65] (see Fig. \ref{powsfig}).
Thus, the amplitude of $\sigma_8$ for a flux-limited I$\leq$24 sample
increases as a function of redshift by nearly $70\%$ between z$\sim$0.3 and z$\sim$1.7.

\subsection{The 3-Dimensional VVDS Density Field} 

The VVDS-02h galaxy density field reconstructed on a scale R= 5 \hpc
and in the redshift bin
0.8$<$\z$<$1.1 is visually displayed in the left most panel 
of Fig. \ref{figco}. Note that the chosen  smoothing
scale nearly corresponds to the mean inter-galaxy separation 
in the VVDS-02h sample at \z$\sim$ 1. Fig. \ref{figco} shows the regular 
patterns traced by over- and under-dense  regions in the selected 
redshift interval. Specifically, we note that, in this redshift slice, 
there are over- and under-dense regions which extend over characteristic scales 
as large as $\sim$ 100 \hpcp
A more complete discussion of the ``cartography" in such  deep regions of the Universe is
presented by \citet{lef05c}.

In the same figure, we also display the density field reconstructed in 
an analogous redshift range, using the GALICS simulation
(Galaxies in Cosmological Simulations, \citet{hat}). GALICS 
combines cosmological simulations of dark
matter with semi-analytic prescriptions for galaxy formation to produce a fully 
realistic deep galaxy sample.
In particular we
plot the density field of the I$\leq$24 flux-limited simulation as well as
the density field recovered after applying to the pure flux-limited
simulation all the VVDS target selection criteria (see \S 4.1). 
No qualitative difference between  the density fields reconstructed before and
after applying to the simulation all the survey systematics is seen.

Clearly, a more quantitative assessment of the robustness 
and reliability of the VVDS overdensity field 
can be  done by studying its PDF.

\section{The PDF of Galaxy Fluctuations}

Once the three-dimensional field of galaxy density contrasts $\delg$ has been
reconstructed on a given scale R, one can fully describe its properties by
deriving the associated PDF $g(\delg)$.
This statistical quantity represents the normalized probability
of having a density fluctuation in the range
$(\delg,\delg+d\delg)$ within a region of characteristic length R
randomly located in the survey volume.

While the shape of the PDF of mass fluctuations at any given cosmic
epoch is theoretically well constrained from CDM simulations (see
next section), little is known about the observational PDF of the
general population of galaxies in the high redshift Universe.
Even locally, this fundamental statistics has been often overlooked
(but see \citet{ost03}). Notwithstanding, 
the shape of the galaxy PDF is strongly sensitive to the effects
of gravitational instability and galaxy biasing,
and its redshift dependence encodes valuable information 
about the origin and evolution of galaxy density fluctuations.

The shape of the PDF can be characterized in terms of its statistical
moments.  In particular the variance of a zero-mean field (such as the
overdensity field we are considering) is

\begin{equation} \sig_{R}^2 = \vev{\delg^2}_{R} \equiv
\int_{0}^{\infty} \delg^2 g_{R}(\delg) d \delg.  \label{var}
\end{equation}

Higher moments can be straightforwardly derived \citep[see][for a
review]{ber02}. In the following, we will drop the suffix R, unless we
need to emphasize it.

\subsection{Estimating Reconstruction Systematics Using Mock Catalogs}

For the purposes of our analysis, it is  imperative to check that the various instrumental 
selection effects as well as  the VVDS observing strategy  are  not compromising
the determination of the PDF of galaxy fluctuations. 
In this section,  we explore the region of the parameter space 
(essentially redshift and smoothing scales) where the  PDF of VVDS 
overdensities traces in a statistically  unbiased way the underlying
parent distribution.

Possible systematics can be hidden in the reconstructed PDF
essentially because i) the VVDS redshift sampling rate is not unity,
ii) the slitlets are allocated on the VIMOS masks with
different constraints along the dispersion and the spatial axis, and
iii) the VIMOS field of view is splitted in four different rectangular
quadrants separated by a vignetting cross.

We have addressed point iii) by designing a specific telescope
pointing strategy which allows us to cover in a uniform way the survey
sky region (see the telescope pointing strategy shown in Fig 1 of Paper I).
With the adopted survey strategy,  we give to each galaxy in the VVDS-02h-4 field
four chances to be targeted by VIMOS, thus  increasing the survey sampling
rate (nearly 1 galaxy over 3  with magnitude I$\leq$24 has a measured
redshift.)

Concerns about points i) and ii) can be directly addressed using
galaxy simulations covering a cosmological volume comparable to the
VVDS one. Thanks to the implementation of the Mock Map Facility
\citep[MoMaF][]{bla03}, it is possible to convert the GALICS 3D mocks catalog
into 2D sky images, and handle the 2D projection of the
simulation as a pseudo-real imaging survey.
Pollo et al. (2005), have then built  a set of 50 fully realistic 
mock VVDs surveys from the GALICS simulations to which the whole observational
pipeline and biases has been applied.
By comparing specific properties of the resulting
pseudo-VVDS sample with the true underlying properties of the
pseudo-real Universe from which the sample is extracted, we can
directly explore the robustness, as well as the limits, of the
particular statistical quantities we are interested in.

In brief these include addeding to the 3D galaxy mocks a randomly simulated
distribution of stars to mimic the same star contamination affecting
our survey.  Next,  we have masked
the sky mocks using the VVDS photometric masks, \ie we have
implemented the same geometrical pattern of excluded regions with
which we avoid to survey sky regions contaminated by the presence of
bright stars or photometric defects.  Then, we have extracted the
spectroscopic targets by applying the target selection code 
\citep[VMMPS,][]{bot04} to the simulated 2D sky distribution.  
To each GALICS redshift, which 
incorporates the Doppler contribution due to galaxy peculiar velocities, we
have added a random component  to take into account errors in z measurements.
Finally,
we have processed the selected objects implementing the same
magnitude-distribution of failures in redshift measurements which
characterizes the first-epoch observations of the VVDS survey
(see Paper I and Fig \ref{figsr}).

Since GALICS galaxies have magnitudes simulated in the same 4 bands
surveyed by VVDS (B,V,R,I), we have applied the K-correction to
obtain rest frame absolute magnitudes and we have empirically
re-derived all the selection functions for the mock catalogs according
to the scheme presented in \S 3. In this way we can also check  the robustness
of the techniques we apply for computing absolute magnitudes and selection
functions (see Paper II).

\begin{figure} \centering \includegraphics[width=9.3cm, angle=0]{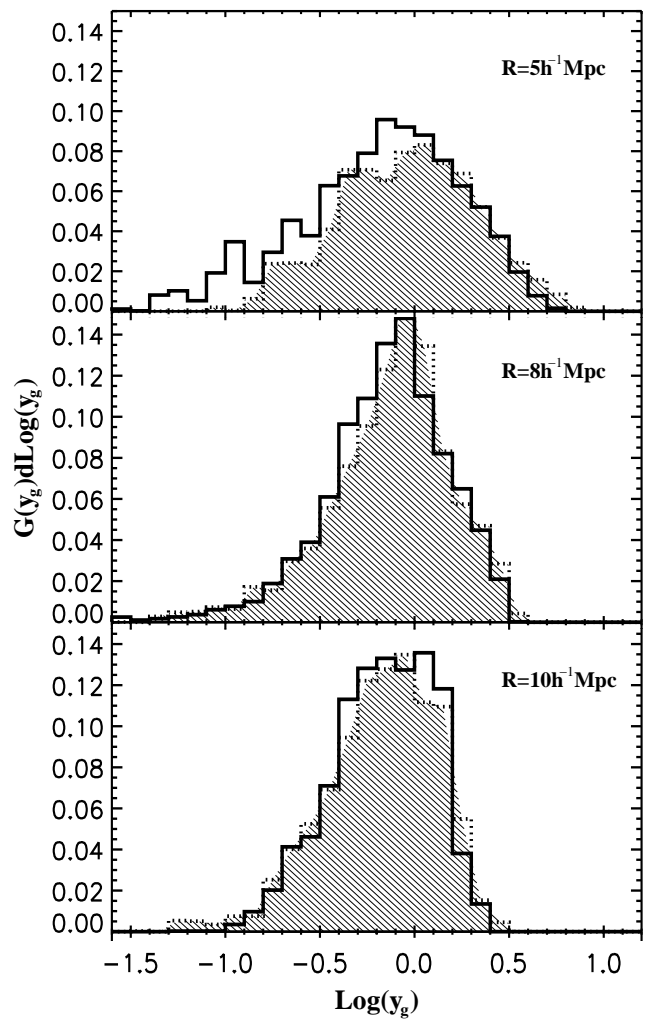}
\caption{Tests of the PDF reconstruction scheme using the mocks VVDS samples
extracted from GALICS.
The differential ($G(y_g)$)  
probability distribution functions  of  $y_g=1+\delg$
for the "observed" s-sample (dotted line, shadowed histogram), and for the 
parent p-sample (solid-line). 
Note that the plotted histograms actually corresponds to 
$G(y_g)=\ln(10)y_gg(y_g))$ since the binning is done in $\log(y_g)$. 
The logarithmic PDFs
are computed for density fields smoothed using TH
filters of different sizes (indicated on the top of each panel).}
\label{figgy} \end{figure}

The PDF of galaxy overdensities obtained from the  mock samples has been finally
compared to the PDF of the parent population. For brevity, in
what follows, we will call {\it s-samples} (survey-samples) the mocks
simulating the VVDS redshift survey and {\it p-sample}
(parent sample) the whole GALICS simulation flux-limited at 17.5$\leq$I$\leq$24.

The density contrasts have been calculated as described in \S 3.  In
the following, we will restrict our analysis to the set of smoothing
scales in the interval R =(5,10) \hpcp The choice of these particular
limits is motivated by the fact that 5 \hpc is the minimum smoothing scale
for which the reconstructed density field is unbiased over a substantial
redshift interval (see discussion below). Note, also, that below this typical
scale, linear regimes approximations, largely used in this paper, do
not hold anymore.  The upper boundary is constrained by the transverse
comoving dimensions covered  by the first-epoch VVDS data (see \S 2),
which is still too small for
being partitioned using bigger scale-lengths without introducing
significant noise in the reconstructed PDF (see the transverse
comoving dimension of the VVDS-02h field quoted in \S 2).

\begin{figure} \centering \includegraphics[width=9cm, angle=0]{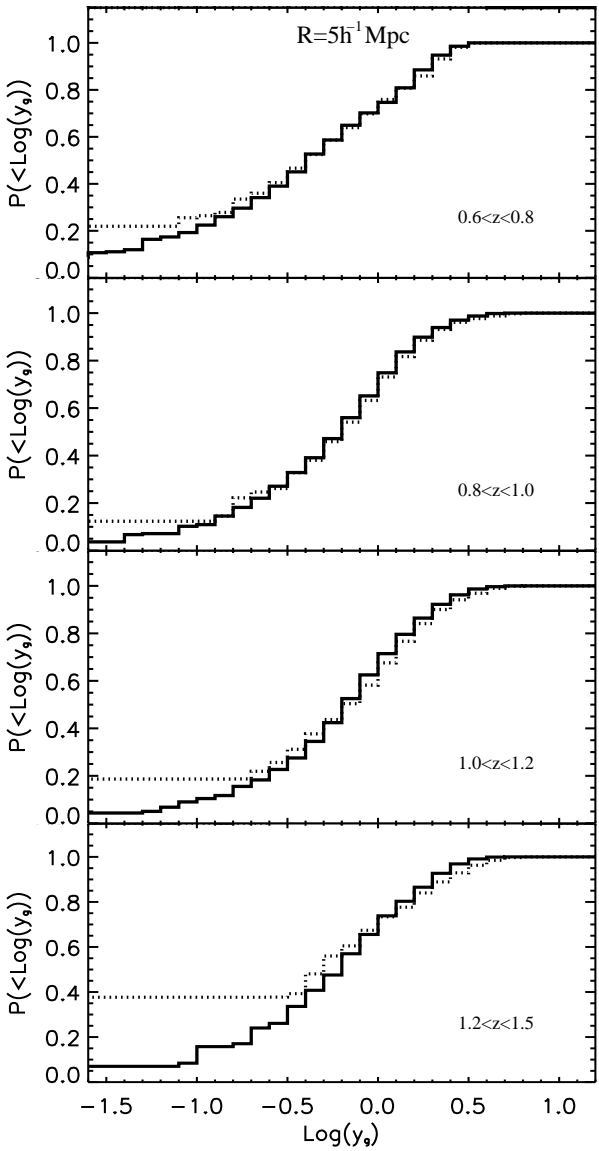}
\caption{ The cumulative distribution function of density contrasts 
$y_g=1+\delg$, on scales R=5 \hpcv  as recovered in different redshift 
intervals for the s-sample (dotted line), and for the 
parent p-sample (solid-line). 
The flattened pedestal 
at the low-density end of the cumulative distribution 
is due to low density regions in the p-sample that
are spuriously sampled as empty regions ($\delg=-1$) in the s-sample.}
\label{figgyb} \end{figure}

The PDF of the galaxy density contrasts computed using the s-sample is
compared to the parent distribution inferred from the p-sample in
Fig. \ref{figgy}. 
 We conclude that the distribution of galaxy
overdensities of the s-samples for $R=8$ and $10$ \hpc scales is not biased
with respect to the underlying distribution of p-sample galaxy
fluctuations. Thus, the VVDS density field reconstructed  
on these scales  is not affected by the specific VVDS observational 
strategy.

It is evident in Fig. \ref{figgy} that the VVDS redshift
sampling rate is not high enough to map in an unbiased way the low
density regions of the Universe ($\log(1+\delg) \lesssim -0.5 $) when the galaxy
distribution is smoothed on scales of  5 \hpcp 
Fig. \ref{figgyb} shows that
incompleteness in underdense regions is a function of redshift, with
the bias in the low-density tail of the PDF developing and increasing
as the redshift increases.
As a rule of thumb,
the  PDF of the s-sample starts to deviate significantly from the parent 
PDF when the mean inter-galactic separation $\vev{r}$ of the 
survey sample is larger than the scale R on which the field is reconstructed. Since 
we measure $\vev{r(z=1)} \sim 5$ \hpc ($\vev{r(z=1.5)}\sim 8$ \hpc),
the PDF of the density field recovered using a TH
filter of radius 5 \hpc is effectively unbiased 
(at least over the density range we are interested in, \ie $log(1+\del)>-1$) 
only if the sample is limited at \z$\lesssim1$. Therefore, in the following,
results obtained for R=5 \hpc are quoted only up to z=1.

On scales R$>8$\hpc, the agreement between the PDFs of s- and p-samples 
holds true also for volume-limited subsamples
Specifically, the 2nd and 3rd moments 
of the PDF of overdensities  recovered using volume-limited {\it s-samples} 
on these scales are within 1$\sigma$ of the corresponding values computed for the 
parent, volume-limited,  {\it p-samples} in each redshift interval of interest up
to \z=1.5.

To summarize, the results of simulated VVDS observations 
presented in this section show that, at least on  scales 
R$\ge$8 \hpcv the VVDS PDF describes in an unbiased way the
general distribution properties of a sample of I=24 flux-limited galaxies up to
redshift 1.5.  In other terms, the VVDS density
field sampled in this way  is essentially free from selection 
systematics in both low- and
high-density regions, and can be  meaningfully used to  
infer the physical bias 
in the distribution between galaxy and matter.
Obviously, the representativeness of the measured  PDF of VVDS overdensities 
with respect to the 
``universal" PDF up to \z=1.5  is a different question. Since
the volume probed is still restricted to a limited region of space
in one field, the shape and moments of the galaxy PDF derived from the VVDS-02h
are expected to deviate from the ``universal" PDF of galaxies at this redshift
because of  cosmic variance.  Reducing the cosmic variance 
is one of the main goals of extending the VVDS to 4 independent fields.
Anyway, our 50 mock realisations
of the VVDS-02h sample allow us to estimate realistic errors that 
include the contribution from cosmic variance.

\subsection{The PDF of VVDS Galaxies: Results}

Let us then investigate the evolution, as a function of
the lookback time, of the observed PDF of VVDS galaxy fluctuations. 

In an apparent magnitude-limited survey such as the VVDS, only
brighter galaxies populate the most distant redshift bins, whereas
fainter galaxies are visible only at 

\begin{figure*} \centering \includegraphics[width=18cm, angle=0]{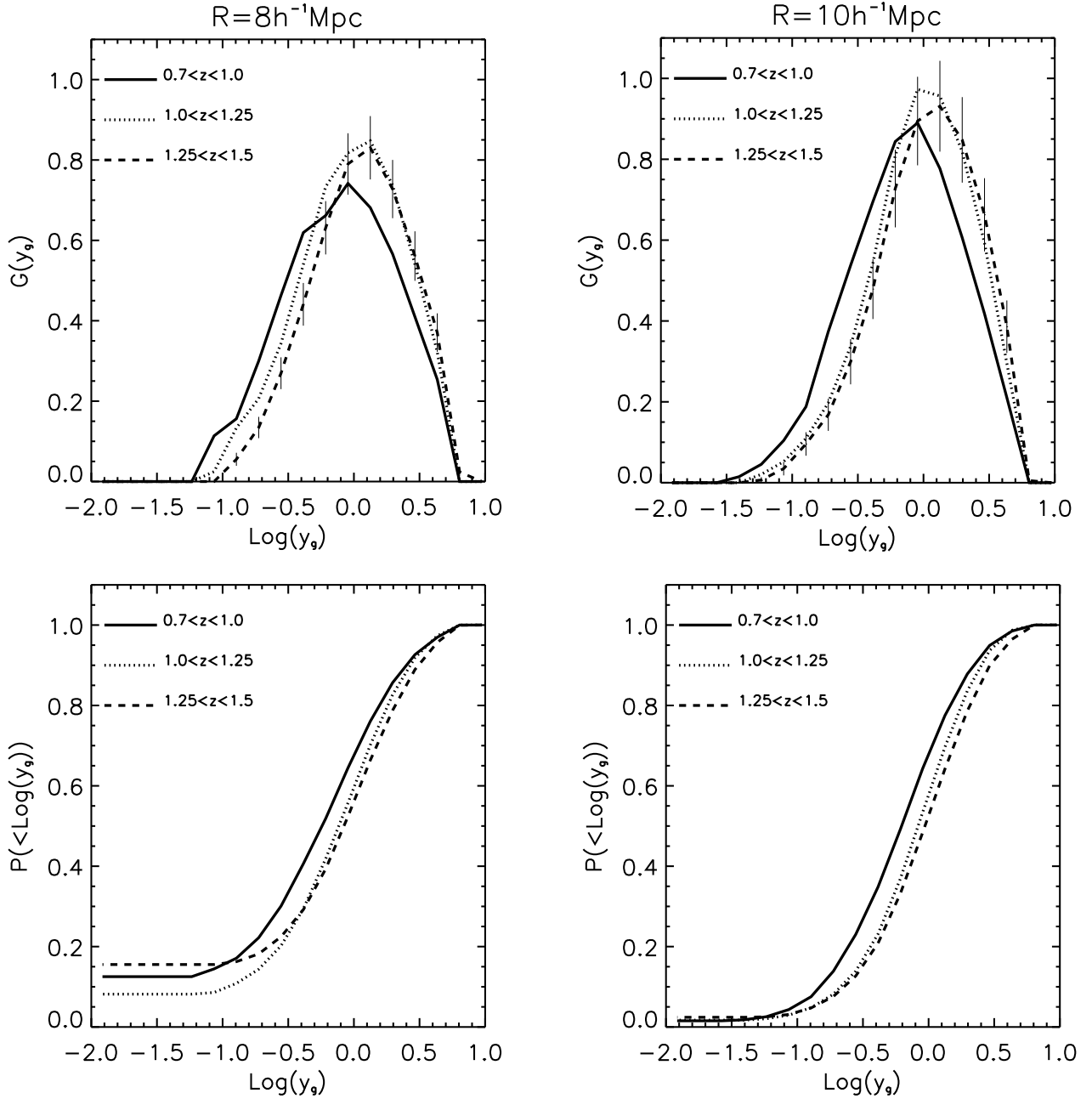}
\caption{{\it Top:} The PDF per units of  galaxy overdensities ($G(y)=\ln (10)yg(y)$) 
is plotted for the volume limited VVDS sample
($\mathcal{M}^{c}_B=-20+5\log h$) at three different redshifts.  The
PDFs  are computed for density fields recovered using TH filters of size 8 and 10 \hpcp
{\it Bottom:} the corresponding cumulative distributions. Errorbars (which, for clarity,
are plotted only for the redshift bin $1.25<z<1.5$)
represent the Poissonian uncertainties. The flattened pedestal 
at the low-density end of the cumulative distribution 
is contributed by empty regions ($\delg=-1$).}
\label{figvu} \end{figure*}

\noindent lower redshifts.  As more luminous
galaxies tend to cluster more strongly than fainter ones \citep[\eg][]{ham88,cro04}, 
the PDF of galaxy fluctuations is expected to be systematically biased as a
function of redshift. 

This effect is clearly seen in the first correlation analysis 
of the VVDS (paper III),  and can be minimized by selecting a
volume-limited sample. Therefore, we have defined a subsample 
with absolute magnitude brighter than $\mathcal{M}^{c}_B=-20+5\log h$ 
in the rest frame B band ($\sim 1350$ galaxies with 0.7$<$\z$<$1.5 
in the VVDS-02h field, $\sim 800$ of which are in the  VVDS-02h-4
region). 

This threshold corresponds to the faintest galaxy luminosity 
visible at redshift \z=1.5 in a I=24 flux-limited survey and 
it is roughly 0.6 magnitudes brighter(fainter) than the 
value of $\mathcal{M}_{B}^*$ recovered  at \z$\sim$0($\sim$1.5)  using
the VVDS data (see Paper II).  The median absolute magnitude for 
this volume-limited sample is $\sim -20.4$. 

We note, however, that the populations of galaxies with the same luminosity at different redshifts may actually
be different. As we have shown  in Paper II, we measure
a substantial degree of evolution in the luminosity of galaxies, and, as a consequence, 
 with our absolute magnitude cut 
we are selecting $\mathcal{M}^{*}+0.6$ galaxies at \z=1.5, but $\mathcal{M}^{*}-0.6$ galaxies at \z=0.
Thus, the clustering signal at progressively earlier epochs may not be  contributed 
by the progenitors of the galaxies that are sampled at later times
in the same luminosity interval.

The PDF of density fluctuations, in various redshift intervals, and  traced on  scales of 8 and 10 \hpc  
by VVDS  galaxies brighter than $\mathcal{M}^{c}_B$, is presented in
Fig. \ref{figvu}. Note that the analysis of the previous section guarantees that, on these scales,  
the VVDS PDF fairly represents the PDF
of the real underlying population of galaxies  up to \z=1.5. 
Fig. \ref{figvu} shows how the
shape of the measured galaxy PDF   changes across different cosmic epochs.

A Kolmogorov-Smirnov test confirms 
that the PDFs at different cosmic epochs are statistically
different (\ie the null hypothesis that the three distributions are drawn 
from the same parent distribution is rejected with a confidence $P_{ks}>1-10^{-6}$).

In particular the peak of the PDF in the lowest redshift interval is shifted 
towards smaller values of the density contrast $\delg$ 
when compared to the peak of the PDF in the highest redshift bin.
Moreover, the shape of the distribution, 
also shows a systematic ``deformation". Specifically, 
we observe the development of a low-$\del$ tail in the PDF as
a function of time on both  scales investigated. In other terms
the probability of having low density 
regions increases as a function of cosmic time. 
For example, underdense  regions, defined as the regions where the galaxy density
field is $\log(1+\delg)<-0.5$ on a R= 8 \hpc scale, occupy a fraction of nearly 
35$\%$ of the 
VVDS volume at redshift 0.7$<$\z$<$1.0, but only a fraction of about 25$\%$ at 
redshift 1.25$<$\z$<$1.5. 
Similar trends are observed  when 
lowering the absolute luminosity threshold of the volume limited sample 
(and consequently lowering the upper limit of the redshift interval probed) 
or when modifying the binning in redshift space.

   If galaxies are faithful and unbiased tracers of the underlying dark matter field, 
   then   
   both this effects, the peak shift and the development of a low density tail 
   may be qualitatively  interpreted  
    as a direct supporting evidence for the  paradigm of the 
   evolution of gravitational clustering in an expanding Universe. 
   At variance with overdense regions which collapse, a net density deficit ($\del<0$)
   in an expanding Universe brings about a sign reversal of the effective 
   gravitational force: a density depression is a region that induces an effective 
   repulsive peculiar gravity (Peebles 1980). If gravity is the engine which drives 
   clustering in an expanding Universe, we thus expect that, as time goes by, low
   density regions propagate outwards and a progressively higher portion  of the 
   cosmological volume becomes underdense.

\begin{figure} \centering \includegraphics[width=9cm, angle=0]{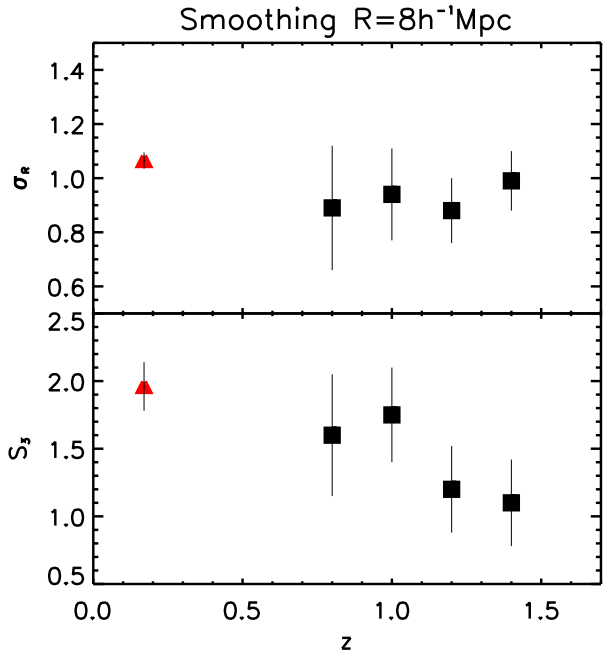}
\caption{Redshift evolution of the standard deviation ({\it upper panel} 
and of the skewness ({\it lower panel}) of the galaxy PDF  on a scale  R=8 \hpc
for galaxies brighter than $\mathcal{M}=-20+5\log h$.
The corresponding local values,  
estimated on the same scale by Croton et al. 2004 
using a subsample of the 2dFGRS having nearly the same median
absolute luminosity of our sample, are represented with triangles.
Error bars represent 1$\sigma$ errors,
and, in the case of VVDS measurements,  include the contribution from cosmic variance.
The errorbar on $\sigma_8$ of the 2dFGRS is smaller than the symbol size.}
\label{figvum} \end{figure}

   The observed evolution in the PDF 
   could  also indicate the presence of a time-dependent biasing between matter 
   and galaxies. 
   As a matter of fact, it can be easily shown that a monotonic bias, increasing 
   with redshift, offers a natural mechanism 
   to re-map the galaxy PDF  into progressively  higher intervals of density contrasts. 
   
   We can better discriminate the physical origin of the observed trends, \ie if they 
   are purely induced by gravitation 
   or strengthened by the collateral and cooperative  action of biasing, by 
   studying the evolution of the PDF moments.
In Fig. \ref{figvum} the redshift evolution of the {\it rms} ($\sigma$) and skewness
($S_3$) of the overdensity fields (see table 1) for the $\mathcal{M}_{B}<-20+5\log h$ sample are shown and compared to local measurements. 

Following the standard convention within the hierarchical clustering 
model, we define the skewness $S_3$ in terms of the volume-averaged
two- and three-point correlation functions ($S_3 \equiv \bar{\xi}_3/\bar{\xi}_2^2$)
noting that in the case of a continuous
$\delg$-field with zero mean this expression reduces to 
$S_3 \equiv \vev{\delg^3}/\vev{\delg^2}^2$.
We do not 
derive the moments $<\delg^2>$ and $<\delg^3>$ of the PDF by directly applying the computation scheme given in eq. (\ref{var}), but by correcting the count-in-cells
statistics for discreteness effects using the Poissonian shot-noise  
model \citep[\eg][cfr. eqs. (374) and (375) of Bernardeau et al. 2002,
possible biases introduced by this estimation technique are discussed by
Hui \& Gazta\~naga 1999.]{pee80,fry85}.
The corresponding values of $\sigma$ and $S_3$ for the local Universe  
(in redshift space)
have been  derived by Croton et al. 2004 using the 2dFGRS. Here we plot 
the values corresponding to their $-21<\mathcal{M}-5\log h<-20$  subsample, which actually brackets the median luminosity of our volume limited sample.

We can see that
the {\it rms} amplitude of  fluctuations 
of the VVDS density field, on scales 8 \hpcv
is with good approximation constant 
over the full redshift baseline investigated, with a mean value 
of 0.94$\pm$0.07
over 0.7$<$\z$<$1.5.

While 
the strength of clustering of galaxies brighter than   $\mathcal{M}<-20+5 \log h$
does not change much in this redshift interval,    
each VVDS measurement is lower 
than the value inferred at  \z$\sim$0 by Croton et al. 2004.
In particular, our mean value is $\sim$10$\%$ smaller than the 2dFGRS 
value and the difference is significant at $\sim 2\sigma$ level.

The skewness $S_3$, which  measures the
tendency of gravitational clustering to create asymmetries between
underdense and overdense regions, decreases as a function of
redshift. We observe a systematic decrement not only  internally 
to the VVDS sample, but also when we compare our measurements 
with the \z=0 estimate.
This trend is caused by the development of the 
low-$\del$ tail in the PDF as
a function of time on both the R=8,10 \hpc scales and reflects the fact 
that the probability of having underdense regions is greater at present epoch  
than it was  at \z$\sim$1.5 (where its measured value is $\sim 2\sigma$ lower.) 

The amplitudes of the {\it rms} and  skewness of galaxy overdensities  
show  an evolutionary trend dissimilar from that predicted in first and second
order perturbation theory for the gravitational growth of dark matter fluctuations 
(see Bernardeau et al. 2002 for a review). According to linear perturbation
theory the amplitude of the {\it rms} of mass fluctuations scales with redshift 
as in eq. (\ref{siglt}) while 
second order perturbation theory predicts that, on the scales where the quasi-linear 
approximation holds, 
the growth rate of $<\delg^3>$ and variance $<\delg^2>^2$ are syncronized
so that  the skewness $S_3$ of an initially Gaussian fields 
should remain constant \citep{pee80, jus93, ber94}.
\footnote{Note that the observed redshift evolution of the skewness is 
just the opposite of what is expected also  
in generic dimensional non-Gaussian models where $S_3$ is predicted to increase 
with redshift.}
Furthermore, in Le F\`evre et al. (2005c)
we show  that  even the general shape of the galaxy PDF  
deviates from a lognormal distribution, 
\ie from the profile in terms  of which the 
mass PDF is generally approximated (see \S 5). 
Therefore, we conclude that the PDF evolution  
is not caused by gravity alone; the redshift scaling of its global shape 
and moments effectively indicates the presence of a time evolving bias.

We can deconvolve the purely gravitational signature and 
investigate properties and
characteristics of the biasing between matter and galaxies
by comparing the galaxy PDF to the corresponding statistics computed for  mass 
fluctuations. 
Thus, we now turn to the problem of deriving  the PDF of mass fluctuations.

\section{The PDF of Mass Fluctuations in
Redshift-Distorted Comoving Coordinates}

The VVDS survey is providing a rich body of redshift data for mapping the
galaxy density field in extended regions of space and over a wide
interval of cosmic epochs.  On the contrary, the direct determination of
the underlying mass density field and its associated PDF is a less
straightforward process.  Nonetheless we may gain insight into the
mass statistics by using simulations and theoretical arguments.

In the standard picture of gravitational instability, the PDF of the
primordial cosmological mass density fluctuations is assumed to obey a
random Gaussian distribution.  Once the density fluctuations reach the
non-linear stage, their PDF significantly deviates from the initial
Gaussian profile and a variety of phenomenological models have been
proposed to describe its shape \citep[\eg][]{sas85,lah93}. In
particular, it is well established in CDM models that when structure
formation has reached the nonlinear regime, the density contrasts in
comoving space $f(\del)$ follow, to a good approximation, a lognormal
distribution \citep{col91,kof94,tay00,kay01}, 

\begin{equation}
f(\del)=\frac{(2 \pi \omega^2)^{-1/2.}}{1+\del} \exp \Big\{ -\frac{
[\ln(1+\del) +\omega^2/2]^2}{2\,\omega^2} \Big\} \label{teopdf}
\end{equation}

This approximation becomes poor in the highly non-linear regime 
\citep[e.g]{ber95,ued96}. The PDF of mass overdensities $f(\del)$ 
is characterized by a
single parameter ($\omega$) that is related to the variance of the
$\del$-field as 

\begin{equation} \omega^2=\ln [1+\vev{\del^2}]
\label{omega}
\end{equation}

At high redshifts, the variance $\sig_{R}$ over sufficiently large
scales R (those explored in this paper) may be easily derived using
the linear theory approximation: 

\begin{equation}
\sig_R(z)=\sig_R(z=0) D(z) \label{siglt}
\end{equation} 

\noindent where D(z) is the linear
growth rate of density fluctuations normalized to unity at \z=0 
\citep{hea77,ham01}.

The lognormal approximation formally describes the distribution of
matter fluctuations computed in real comoving coordinates. On the contrary,
the PDF of galaxies is observationally derived in redshift space.  In
order to map properly the mass overdensities into galaxy overdensities
the mass and galaxy PDFs must be computed in a common reference frame.
It has been shown by \citet{sbd00} that an optimal strategy to derive
galaxy biasing is to compare both mass and galaxy density fields
directly in redshift space.  Implicit in this approach is the
assumption that mass and galaxies are statistically affected in the
same way by gravitational perturbations, and thus, that there is no
velocity bias in the motion of the two components.

A general model which allows the explicit computation of the statistical
distortions caused by peculiar velocities has been proposed by Kaiser
(1987).  This applies in the linear regime (\ie on large scales)  and in the local Universe where
redshift and distances are linearly related.  At cosmological
distances z, however, the mapping between real comoving coordinates
($\blx$) and redshift comoving coordinates ($\bly$), \ie the
pseudo-comoving coordinates inferred on the basis of the observed
redshifts, is less trivial, and we proceed to obtain it in the following.

In an inhomogeneous Universe, galaxies have motions above and beyond
their Hubble velocity \citep[\eg][]{gio,mar98,bra}.  As a consequence, Doppler 
spectral shifts add to the cosmological signal and the observed redshift
$(\tilde{z})$ is given by

\begin{equation} \tilde{z}=z+\frac{U(\blx)}{c}(1+z) \end{equation}

\noindent where \z$\,$ is the cosmological redshift in a uniform
Friedman-Robertson-Walker metric and where $U(\blx)={\bf v(\blx)\cdot
\hat{r}}=|{\bf v(\blx)}|\mu$ is the radial component of the peculiar velocity
($\mu$ is the cosine of the angle between the peculiar velocity vector
and the line-of-sight versor ${\bf \hat{r}}$).

The redshift comoving distance of a galaxy at the observed redshift $\tilde{z}$ is 
thus

\begin{equation} y=\frac{c}{H_0}
\int_{0}^{z+\frac{U}{c}(1+z)}\frac{1}{E(\chi)}d\chi,  
\end{equation}

\noindent where

\begin{equation}
E(z)=[\Omega_m(1+z)^3+(1-\Omega_m-\Omega_{\Lambda})(1+z)^2+\Omega_{\Lambda}]^{1/2}.
\end{equation}

At high redshifts ($z \gg \frac{|U|}{c}$),  we can write

\begin{equation} y=\frac{c}{H_0} \Bigg[
\int_{0}^{z}\frac{1}{E(\chi)}d\chi +\frac{U}{c}(1+z)E(z)^{-1}\Bigg]
\end{equation}

\noindent which, in turns, gives the coordinate transformation
 from real comoving space $\blx$ to the redshift comoving space $\bly$

\begin{equation} \bly=\blx \Bigg[1+p(z)\frac{U(\blx)}{\blx} \Bigg].
\end{equation}

In this mapping, the cosmological term $p(z)$, 

\begin{equation}
p(z)=\frac{1+z}{H_0 E(z)} 
\end{equation} 

\noindent is a correcting factor which
takes into account the fact that, at high redshifts, distances do not
scale linearly with redshift, and, thus, that  peculiar velocities cannot be
simply added to redshift space positions as in the local Universe.

The galaxy density field in the redshift-distorted space is related to
the galaxy density in real space by the Jacobian of the transformation
between the two coordinate systems

\begin{equation} \rho_y(\bly)= \rho_x(\blx)
\Big[1+p(z)\frac{U(\blx)}{\blx}\Big]^{-2}\Big[1+p(z)\frac{dU(\blx)}{d\blx}\Big]^{-1}
\end{equation}

At sufficiently large distances from the observer, neglecting the
survey selection function (\ie  considering a volume-limited redshift
survey) and at first order in perturbations we obtain

\begin{equation} 
\del_y(\bly)= \del_x(\blx) - p(z)\frac{dU(\blx)}{d\blx}. 
\end{equation}

The second term on the right hand side can be evaluated using linear-regime
approximations and gravitational instability theory. In comoving
coordinates it is given by

\begin{equation} 
\frac{dU(\blx)}{d\blx}=-\frac{\mu^2 \rm{f}(z) H(z)}{1+z} \del_x(\blx) 
\end{equation}

\begin{figure} 
\centering 
\includegraphics[width=8cm]{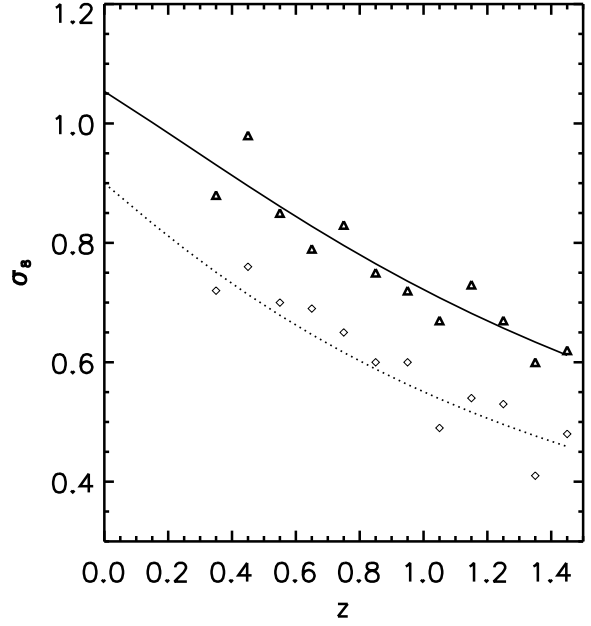}
\caption{Redshift scaling of the {\it rms} mass fluctuations in sphere
of 8 \hpc radius.  Diamonds represent $\sig_8$ as computed from the
$\Lambda$CDM Hubble volume simulation in real comoving space
(x-space), while triangles represent the corresponding values recovered
in the redshift perturbed  comoving coordinates
(y-space). The solid line is the analytical prediction for the scaling
of $\sig_8$ in the y-space obtained using eq. (\ref{sigcor}), while the
dotted line represents the x-space evolution predicted in real
space. In both cases the power spectrum of perturbations is the same
and has been normalized in order to match the simulation
specifications ($\sig^{x}(z=0)=0.9$).}  
\label{figs8}
\end{figure}

\noindent where $\rm{f}=d \ln D/d \ln a$ is the logarithmic derivative of
the linear growth rate of density fluctuations with respect to the
expansion factor a(t). At redshift $z$ (corresponding to the comoving position \blx)
a useful approximation is given by: 

\begin{equation} \rm{f}(z) \sim \Omega_m^{3/5} E(z)^{-6/5}(1+z)^{9/5}
\end{equation} 

\noindent (see \citet{mar91, lah91}).

By combining the previous results we obtain

\begin{equation} \del_y(\bly)=\del_x(\blx)[1+\mu^2 \;\rm{f}(z)].
\end{equation}

\noindent which reduces to the \citet{kai87} correction when \z=0

The relation between the azimuthally averaged variances measured in
real and redshift comoving space is

\begin{equation} \sig^y
(z)=\Big[1+\frac{2}{3}\;\rm{f}(z)+\frac{1}{5}\;\rm{f}^2(z)\Big]^{1/2} \sig^x (z).
\label{sigcor} \end{equation}

\begin{figure} \centering \includegraphics[width=9cm]{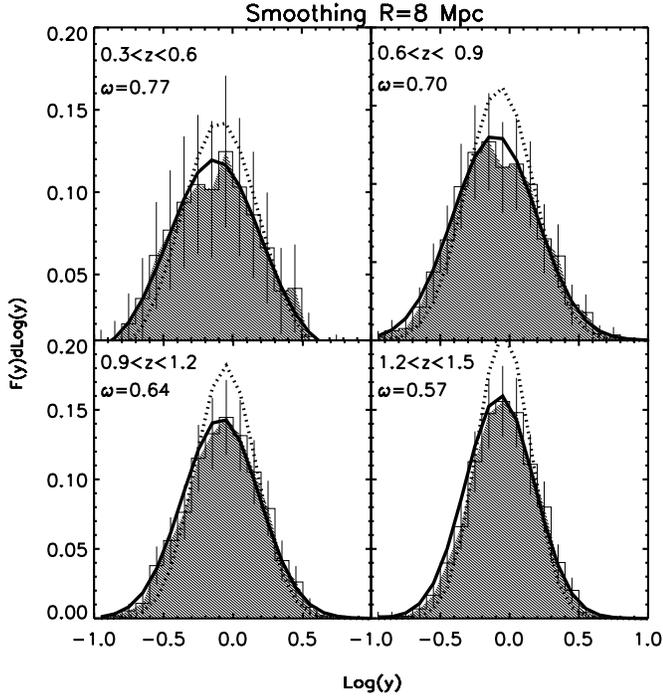}
\caption{One-point PDFs of dark matter fluctuations (shaded area) computed using the
Hubble volume $\Lambda$CDM cosmological simulation in 4 different
redshift ranges over a volume which mimics the geometry of the VVDS
sample.  The mass PDF  has been recovered 
in the redshift comoving space 
by smoothing the mass-particle distribution with  a TH window of size 
R=8 \hpcp  Note that the plotted
histogram actually corresponds to $F(y)=(ln 10)yf(y)$ because the
binning (d$\log y$=0.1) is done in $\log(y)=\log(1+\del)$.
The dotted line represents the lognormal approximation
derived in the real comoving space using eq. (\ref{teopdf}).
The solid curve represents the lognormal approximation computed
adopting the variance parameter $\omega$ (shown in the inset) theoretically inferred using
equation (\ref{sigcor}), which models peculiar velocity distortions as a function
of redshift.} \label{figfy}
\end{figure}

We have tested the validity of eq. (\ref{sigcor}) in the high redshift
domain using the Hubble volume N-body simulations carried out by the
Virgo consortium \citep{col00}.  This is a large numerical experiment which allows
the simulation of mass surveys along the observer past light cone.
The simulated mass distribution is computed in a $\Lambda$CDM
cosmogony with parameters $\Omega_m=0.3$, $\Omega_{\Lambda}=0.7$ and
$H_0=70$ km s$^{-1}$Mpc$^{-1}$.  The volume covered by this N-body
simulation is large enough that the mass survey extracted along the
diagonal of the simulation cubes extends up to the redshift of
interest \ie the redshift covered by the first-epoch VVDS data
(\z=1.5). In this simulation the mass-particle resolution is $2.2\cdot
10^{12}h^{-1}$M$_{\odot}$  and the present epoch is defined by a linear {\it rms}
density fluctuation in a sphere of radius 8 \hpc of $\sig_8=0.9$.

The mass density contrasts in the redshift perturbed comoving
coordinates $\del(\bly_i, R)$ have been calculated at random positions
$\bly_i$ in the simulation volume,
by smoothing the particle distribution with a spherical top hat window
of length $R=8$ \hpcp  Mass variances in different redshift bins are then
derived using eq. (\ref{var}). The result is compared to the prediction of
eq. (\ref{sigcor}) in Fig. \ref{figs8}.  Note that, even if it is clear
that measurements suffer from cosmic variance due to the relatively
small volume sampled at each redshift, the predictions of
eq. (\ref{sigcor}) are in agreement with the observed scaling of the
linear mass variance. The magnitude of the
correction with respect to the unperturbed case is also evident; mass fluctuations
recovered in redshift space on a 8 \hpc scale, in the redshift comoving coordinates, at \z=0.5(1.5) are
$\sim 25(35)\%$ larger than in real comoving space (the correction
factor is $\sim 17\%$ in the local Universe.)
Fig. \ref{figfy} shows that this apparent  enhancement in the ${\it rms}$ fluctuations 
results in a  broadening of the mass PDF recovered in the redshift comoving space.
Thus, the effect of peculiar velocities is to shrink overdense regions and to inflate
underdense regions, enhancing the probability of having large 
density fluctuations (both positive and negative).

We finally  compare, in various redshift intervals,  
the accuracy with which  the lognormal mass PDF derived in the redshift comoving space  
(by using  eq. (\ref{sigcor}) in \ref{omega}) approximates the PDF 
directly inferred from the Hubble volume simulation (see Fig. \ref{figfy}). 
On a scale of 8 \hpcv the agreement between
the analytical and simulated mass  PDFs
is satisfactory at all redshifts. This  holds true also when the mass PDFs 
recovered on $R=5$ and 10 \hpc scales  are compared.

Thus, with a good degree of confidence, we can use eq. (\ref{sigcor})
to predict the PDF of mass fluctuations in redshift distorted 
comoving coordinates (the same coordinates where the galaxy PDF is
observed) and in a generic cosmological background. This allows us to
speed up computation time and to frame the results about the biasing
function in a generic cosmological model.

\section{Measuring Galaxy Biasing}

In this section we describe the method applied to determine the
relationship between galaxy and mass overdensities.
The galaxy overdensity field $\delg$ depends in principle on various
astrophysical and cosmological parameters such as spatial position
(r), underlying matter density fluctuations  ($\del$), scale R with which the
density field is reconstructed, cosmological time (z), galaxy colors, local gas
temperature, non-local environment, etc.  

For the purposes of this study, we will rely on the following
simplifying theoretical assumptions:\\
i) the efficiency of galaxy formation on a given
   cosmological scale   is sensitive only to the 
   underlying mass distribution. This means that
the galaxy fluctuation field is in a reasonably tight one-to-one
relationship with the underlying mass fluctuation field, and that the biasing 
scheme may be formally represented via the relationship  $\delg=b(z,\del, R) \del$. 
While such an approach represents a non-trivial step forward in understanding the
properties of the biasing function b (if compared, for example, to
constant parameterizations of the biasing relation), it is however
evident that the biasing function could show, in principle, a more
complex functional dependence. \\ 
ii) The current theoretical
understanding of how clustering of DM proceeds via gravitational
instability in the expanding Universe is well developed, \ie  the PDF
of mass fluctuations of the real Universe can be safely derived via
analytical models or N-body simulations (see discussion in \S 6)
In particular, in what follows, we will consider a $\Lambda$CDM background 
mass distribution locally normalized to $\sigma_8(z=0)=0.9$.\\
iii) The redshift distortions affect the densities of galaxies and
mass in a similar way, \ie there is no velocity bias between these two
components, and galaxies follow the matter flow.

\subsection{The Method}

As described in \S 1, we derive the relationship between 
galaxy and mass overdensities in redshift space $\delg=\delg(\del)$ as the 
one-to-one transformation which maps the theoretical mass PDF $f(\del)$ into 
the observed  galaxy PDF $g(\delg)$.
A similar method to derive the
biasing function has been proposed  and tested using CDM simulations
by \citet{sbd00} (see also \citet{sza}). This same technique  
has been recently applied in different contexts 
by \citet{mar02}  to derive the mass-to-light ($M=M(L)$) and the
X-ray-to-optical ($L_x=L_x(L)$) functions for a wide mass range of 
virialized systems, and by   \citet{ost03}
to explore the void phenomena in the context of hydrodynamic
simulations.

Using eq. \ref{lbs},\ref{nlbp} and \ref{probc}, we obtain 
the biasing function $b(\del)$ as the solution of the following 
differential equation

\begin{equation} \left\{ \begin{array}{l} \delg(-1)=-1 \\ \\
b^{'}(\del)\del+b(\del)=f(\del) g(\delg)^{-1} \end{array}
\right.  \label{de} \end{equation}

\begin{figure*} \centering
\includegraphics[width=18cm, angle=0]{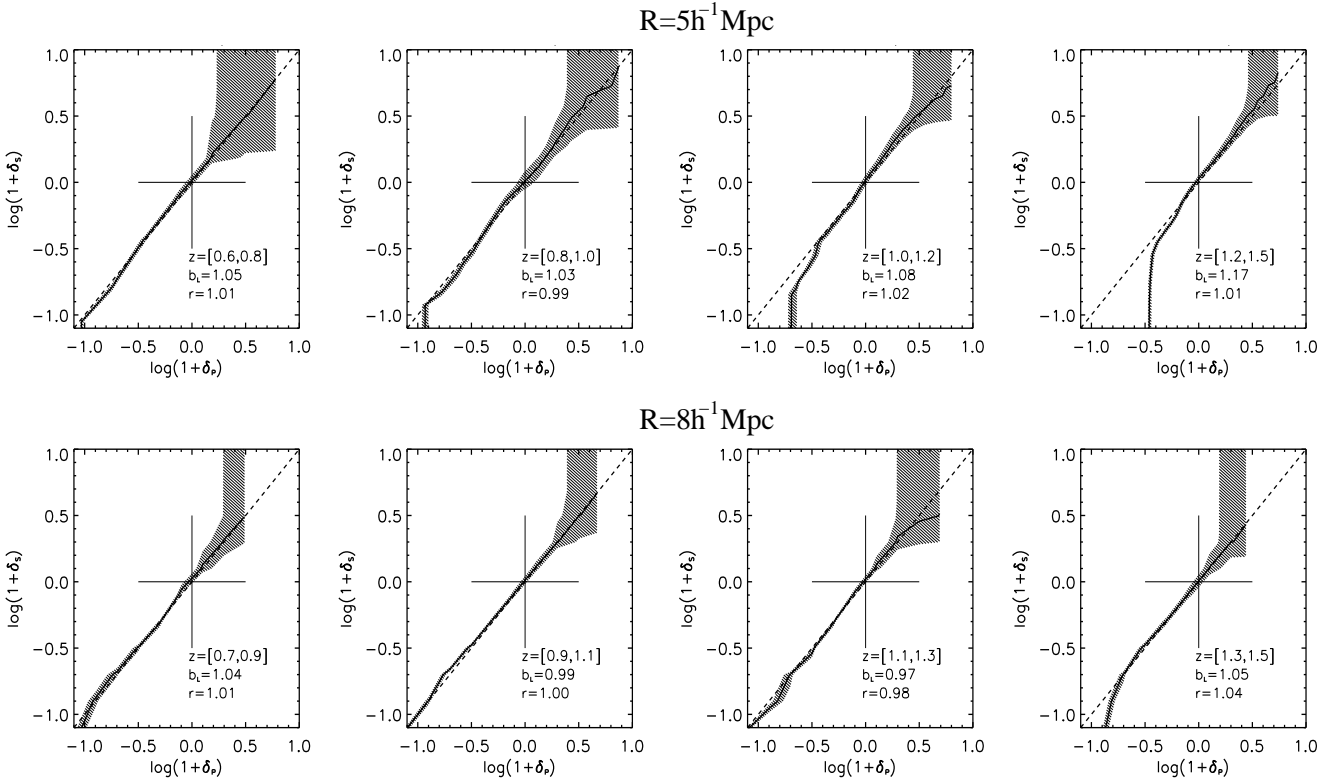} \caption{The simulated biasing
function (solid-line) at different cosmic epochs, between the density
field traced by the s-sample (GALICS data simulating the VVDS sample,
see \citet{pol04} and \S 4.1) and the density field traced by the p-sample (GALICS data
simulating the real underlying distribution of galaxies).  $\bl$
represent the linear bias parameter evaluated from the nonlinear
biasing function using the estimator given in eq. (\ref{lbe}). The
dashed line is drawn at $\bl=1$ and represents the no bias case.
The central cross is for reference and represents the $\delg=\del=0$ case.  The
r parameter measures the deviations from the linearity.  The galaxy
overdensities are reconstructed using a TH window of sizes R=5 \hpc ({\it upper panel})
and R=8 \hpc ({\it lower panel}).
The  shadowed area  represents 1$\sigma$ errors in the derived biasing function.} 
\label{figbls12} \end{figure*}

\noindent where the prime denotes the derivative with respect to
$\del$, $f(\del)$ and $g(\del)$ are the PDF of mass and galaxy
fluctuations respectively, and the initial condition has been
physically specified by requiring that galaxies cannot form where
there is no mass.

With this approach, we loose information on a possible 
stochasticity characterizing the biasing function.
The advantage is that we can provide a
measure, on some characteristic scales R, of the {\it
local}, {\it non-linear}, {\it deterministic} biasing function
(eq. (\ref{nlbp})) over the continuous redshift interval 0.4$<$\z$<$1.5.

We have obtained the biasing function $b(\del)$ by numerically
integrating the differential equation (\ref{de}), i) in different
redshift intervals in order to follow the evolution of $b(\del)$ as a
function of cosmic time, and ii) using matter and galaxy PDFs
obtained by smoothing the density fields on R=5,8, and 10
 \hpc in order to test the scale dependence of the galaxy biasing
function.

The information contained in the non-linear function $b(\del)$ can be
compressed into a single scalar which may be easily compared to the
constant values in term of which the biasing relation is usually
parameterized (see eq \ref{lbs}).  Since, by definition, $\vev{b(\del)
\del}=0$, the most interesting linear bias estimators are associated
to the second order moments of the PDFs,
\ie the variance $\vev{\delg^2}$ and the covariance
$\vev{\delg\,\del}$.  Following the prescriptions of \citet{del99}, we
characterize the biasing function as follows:

\begin{equation} \hat{b} \equiv \frac{\vev{b(\del) \, \del^2}}{
\vev{\del^2}} \label{lbe} \end{equation} 

\noindent and

\begin{equation} \bl^2 \equiv \frac{\vev{b^2(\del)\, \del^2}}{\vev{\del^2}} 
\end{equation}

\noindent where the parameter $\hat{b}$, measuring the slope of the linear
regression of $\delg$ on $\del$, is the natural generalization of the
linear bias parameter defined in equation \ref{lbs} and 
$\bl^2$ is an ``unbiased estimator" of the linear biasing parameter
defined as $\xi_g=b^2 \xi$, when the bias relation is
deterministic, \ie non-stochastic.  The ratio $r=\hat{b}/bl$ is the
relevant measure of nonlinearity in the biasing relation; it is unity
for linear biasing, and it is either larger or smaller than unity for
nonlinear biasing.

The errors in the measured values of the biasing parameter  $\bl$ have
been computed using independent mock catalogs which implement
all the selection functions of the VVDS. This allows us to incorporate in 
our error estimates the uncertainties due to cosmic variance.

\subsection{Testing the Method}

Before applying the biasing computation scheme (eq. \ref{de}) to VVDS
data, we have tested that the method can be meaningfully applied, \ie 
it is free of systematics when implemented with  samples of
simulated galaxies which mimic all the observational systematics of
our sample.

The procedure consists in computing the biasing function
$\del_s=b(\del_p)\del_p$ between the density field $\del_s$
reconstructed using an s-sample (representing the pseudo-survey
sample, see \S 4.1) and the density fluctuations $\del_p$ of the
corresponding p-sample (representing the pseudo-real Universe). 
We have already determined the range of redshift, density contrasts and smoothing scales 
where the   sample simulating 
all the VVDS selection functions (s-sample) 
trace the underlying density of galaxies (p-sample). 
We thus expect, for consistency,
that, in that range, the biasing between the two samples
derived  by applying our computation scheme  (eq. \ref{de}, using the PDFs of the s- and p-samples)  is independent of $\delta_p$ and equal to $b(\del_p)=1$.

Results are presented in Fig. \ref{figbls12}  for
two different TH smoothing scales. Note that a log-log density plot is 
used in order to emphasize the behavior 
of the biasing function in underdense regions.
We conclude
that on scales R$\ge$8  \hpc the density recovered by a ``four-passes" VVDS-like
survey is not biased with respect to the underlying distribution on
any density scale and in any redshift interval up to \z=1.5. As a
matter of fact, the linear bias parameter with which information
contained in the biasing function can be at first order approximated
is $\bl \sim $ 1 and the biasing relation does not show any
significant deviation from linearity as indicated by the fact that the
r parameter is also very close to unity.

If the density field is smoothed on 5 \hpc, the effects of the
incompleteness in low-density regions (already discussed in \S 4.1,
see Figs. \ref{figgy} and \ref{figgyb}) become evident. Underdense regions 
($log(1+\delta_p) \lesssim -0.5$) in volumes at redshift greater
than 1 are poorly sampled with the VVDS survey strategy.

In the same spirit, we have also solved eq. (\ref{de}) for determining
the biasing relation between the PDF of the Hubble volume mass
fluctuations and the lognormal approximation given in
eq. (\ref{teopdf}).  The biasing relation between these two different
descriptions of the mass density field is linear and consistent with
the no-bias hypothesis between the two representations of the density field
on the scales we are interested in ($R\ge5$ \hpc and $\log(1+\del)>-1$).

\begin{table*} 
\begin{center} 
\begin{tabular}{l l l r l l l l}
\hline 
R & Redshift Range & $\mathcal{M}^{c}_B$& N$_{gal}$ & $\bl$  & r & $\sigma_{R}$&$S_{3}$ \\ 
\hpc        & &  & & & &  \\
\hline

5 &0.4$<$\z$<$0.7 &No     &1583   &0.87$\pm$ 0.15 & 0.95  & 0.94 $\pm$0.15 & 1.2$\pm$0.3  \\  
&  0.7$<$\z$<$0.9 &       &1044   &0.95$\pm$ 0.15 & 0.96  & 0.98 $\pm$0.15 & 1.4$\pm$0.3  \\
&  0.9$<$\z$<$1.1 &       &759   &0.97$\pm$ 0.13 & 0.97  & 0.93 $\pm$0.13 & 1.1$\pm$0.3  \\
\hline 			       
			       
5& 0.4$<$\z$<$0.7 &-18.7  &610   &1.06$\pm$ 0.17 & 0.97  & 1.18 $\pm$0.17 & 1.6$\pm$0.3  \\
&  0.7$<$\z$<$0.9 &       &726   &1.03$\pm$ 0.15 & 0.97  & 1.05 $\pm$0.15 & 1.6$\pm$0.3  \\ 
&  0.9$<$\z$<$1.1 &       &751   &1.00$\pm$ 0.14 & 0.97  & 0.97 $\pm$0.14 & 1.5$\pm$0.3  \\ 
 \hline 		       
			       
5& 0.4$<$\z$<$0.7 &-20    &160   &1.10$\pm$ 0.18 & 0.97  & 1.28 $\pm$0.18 & 1.4$\pm$0.4  \\
&  0.7$<$\z$<$0.9 &       &229   &1.12$\pm$ 0.17 & 0.99  & 1.18 $\pm$0.17 & 1.1$\pm$0.3  \\ 
&  0.9$<$\z$<$1.1 &       &289   &1.18$\pm$ 0.15 & 0.97  & 1.17 $\pm$0.15 & 1.4$\pm$0.4  \\

\hline 			       
			       
8& 0.7$<$\z$<$0.9 &No     &1263  &0.92$\pm$ 0.20 & 0.97  & 0.67 $\pm$0.20 & 1.6$\pm$0.4  \\
&  0.9$<$\z$<$1.1 &       &864   &1.03$\pm$ 0.16 & 0.96  & 0.74 $\pm$0.16 & 1.6$\pm$0.3  \\
&  1.1$<$\z$<$1.3 &       &440   &1.21$\pm$ 0.12 & 0.96  & 0.82 $\pm$0.12 & 1.3$\pm$0.3  \\
&  1.3$<$\z$<$1.5 &       &234   &1.51$\pm$ 0.10 & 0.95  & 0.96 $\pm$0.11 & 1.0$\pm$0.3  \\
			       
\hline 			       
			       
8& 0.7$<$\z$<$0.9 &-18.7  &879   &0.98$\pm$ 0.21 & 0.98  & 0.75 $\pm$0.21 & 1.6$\pm$0.4  \\
&  0.9$<$\z$<$1.1 &       &813   &1.01$\pm$ 0.16 & 0.97  & 0.72 $\pm$0.16 & 1.8$\pm$0.3  \\
			       
\hline			       
			       
8& 0.7$<$\z$<$0.9 &-20    &279   &1.17$\pm$ 0.23 & 0.98  & 0.89 $\pm$0.23 & 1.7$\pm$0.4  \\
&  0.9$<$\z$<$1.1 &       &327   &1.30$\pm$ 0.17 & 0.99  & 0.94 $\pm$0.17 & 1.8$\pm$0.3  \\
&  1.1$<$\z$<$1.3 &       &251   &1.33$\pm$ 0.12 & 0.96  & 0.88 $\pm$0.12 & 1.2$\pm$0.3  \\
&  1.3$<$\z$<$1.5 &       &169   &1.56$\pm$ 0.11 & 0.95  & 0.99 $\pm$0.11 & 1.1$\pm$0.3  \\
			      
\hline			      
			      
10&0.7$<$\z$<$0.9 &No     &1425   &1.03$\pm$ 0.22 & 0.91 & 0.66 $\pm$0.22 & 1.6$\pm$0.4  \\
&  0.9$<$\z$<$1.1 &       &955   &1.05$\pm$ 0.18 & 0.97  & 0.64 $\pm$0.18 & 1.7$\pm$0.4  \\
&  1.1$<$\z$<$1.3 &       &480   &1.17$\pm$ 0.13 & 0.90  & 0.68 $\pm$0.13 & 1.3$\pm$0.3  \\
&  1.3$<$\z$<$1.5 &       &250   &1.55$\pm$ 0.11 & 0.93  & 0.84 $\pm$0.13 & 1.2$\pm$0.3  \\
			      
\hline			      
			      
10&0.7$<$\z$<$0.9 &-18.7  & 991  &1.03$\pm$ 0.25 & 0.95  & 0.69 $\pm$0.25 & 1.5$\pm$0.4  \\
&  0.9$<$\z$<$1.1 &       & 900  &1.03$\pm$ 0.18 & 0.96  & 0.63 $\pm$0.18 & 1.7$\pm$0.3  \\
			      
\hline			      
			      
10&0.7$<$\z$<$0.9 &-20    &316   &1.14$\pm$ 0.25 & 0.92  & 0.75 $\pm$0.25 & 1.6$\pm$0.4  \\
&  0.9$<$\z$<$1.1 &       &360   &1.26$\pm$ 0.20 & 0.97  & 0.78 $\pm$0.20 & 1.8$\pm$0.4  \\
&  1.1$<$\z$<$1.3 &       &266   &1.36$\pm$ 0.14 & 0.91  & 0.78 $\pm$0.14 & 1.3$\pm$0.3  \\
&  1.3$<$\z$<$1.5 &       &175   &1.54$\pm$ 0.13 & 0.93  & 0.84 $\pm$0.13 & 1.3$\pm$0.3  \\

\hline 

\end{tabular} 
\caption{Bias measurements from the VVDS first epoch data.}  
\label{table1} 
\end{center}
\end{table*}

\begin{table*} 
\begin{center} 
\begin{tabular}{l c c c c c c c c c}
\hline 
R & Redshift & $\mathcal{M}^{c}_B$& $a_0$ & $a_1$ & $a_2$ & $a_3$ & $b_0$ & $b_1$ & $b_2$ \\ 
\hpc        &range &  & & & & & & & \\
\hline

8& 0.7$<$\z$<$0.9 &No      &$ 0.36\pm0.03$ &$ 1.36\pm0.11$ &$ 0.60\pm0.12$ &$ -0.01\pm0.02$ &$ 0.10\pm0.06$ &$ 0.88\pm0.10$ &$ -0.12\pm0.08$   \\
&  0.9$<$\z$<$1.1 &        &$ 0.35\pm0.03$ &$ 1.36\pm0.40$ &$ 1.08\pm0.10$ &$ -0.08\pm0.02$ &$ 0.23\pm0.06$ &$ 1.18\pm0.09$ &$ -0.20\pm0.06$  \\
&  1.1$<$\z$<$1.3 &        &$ 0.39\pm0.04$ &$ 1.64\pm0.13$ &$ 1.14\pm0.11$ &$ -0.08\pm0.02$ &$ 0.24\pm0.07$ &$ 1.26\pm0.10$ &$ -0.22\pm0.06$  \\
&  1.3$<$\z$<$1.5 &        &$ 0.54\pm0.05$ &$ 2.34\pm0.30$ &$ 1.46\pm0.14$ &$ -0.11\pm0.02$ &$ 0.37\pm0.12$ &$ 1.57\pm0.15$ &$ -0.24\pm0.08$  \\
\hline 			 

8& 0.7$<$\z$<$0.9 &-20     &$ 0.25\pm0.05$ &$ 1.67\pm0.20$ &$ 1.25\pm0.16$ &$ -0.09\pm0.03$ &$ 0.18\pm0.10$ &$ 1.30\pm0.15$ &$ -0.20\pm0.08$   \\
&  0.9$<$\z$<$1.1 &        &$ 0.40\pm0.06$ &$ 2.12\pm0.33$ &$ 1.20\pm0.16$ &$ -0.05\pm0.02$ &$ 0.23\pm0.17$ &$ 1.29\pm0.16$ &$ -0.12\pm0.08$  \\
&  1.1$<$\z$<$1.3 &        &$ 0.45\pm0.04$ &$ 1.86\pm0.13$ &$ 1.29\pm0.13$ &$ -0.11\pm0.02$ &$ 0.31\pm0.08$ &$ 1.40\pm0.13$ &$ -0.26\pm0.06$  \\
&  1.3$<$\z$<$1.5 &        &$ 0.48\pm0.05$ &$ 2.51\pm0.30$ &$ 1.46\pm0.16$ &$ -0.10\pm0.02$ &$ 0.33\pm0.14$ &$ 1.55\pm0.15$ &$ -0.22\pm0.08$  \\
\hline	

10& 0.7$<$\z$<$0.9 &No     &$ 0.25\pm0.03$ &$ 0.85\pm0.10$ &$ 1.44\pm0.21$ &$ -0.23\pm0.06$ &$ 0.17\pm0.07$ &$ 1.20\pm0.15$ &$ -0.30\pm0.12$  \\
&  0.9$<$\z$<$1.1 &        &$ 0.18\pm0.02$ &$ 1.20\pm0.08$ &$ 1.23\pm0.13$ &$ -0.18\pm0.04$ &$ 0.20\pm0.04$ &$ 1.19\pm0.09$ &$ -0.34\pm0.08$  \\
&  1.1$<$\z$<$1.3 &        &$ 0.39\pm0.04$ &$ 1.30\pm0.11$ &$ 1.04\pm0.13$ &$ -0.08\pm0.03$ &$ 0.26\pm0.06$ &$ 1.18\pm0.10$ &$ -0.22\pm0.08$  \\
&  1.3$<$\z$<$1.5 &        &$ 0.54\pm0.04$ &$ 2.09\pm0.13$ &$ 1.41\pm0.15$ &$ -0.14\pm0.03$ &$ 0.39\pm0.08$ &$ 1.55\pm0.12$ &$ -0.32\pm0.08$  \\
\hline

10&0.7$<$\z$<$0.9 &-20     &$ 0.21\pm0.03$ &$ 1.33\pm0.10$ &$ 1.45\pm0.16$ &$ -0.22\pm0.04$ &$ 0.18\pm0.06$ &$ 1.26\pm0.11$ &$ -0.28\pm0.10$  \\
&  0.9$<$\z$<$1.1 &        &$ 0.18\pm0.04$ &$ 1.51\pm0.15$ &$ 1.38\pm0.16$ &$ -0.13\pm0.03$ &$ 0.20\pm0.08$ &$ 1.36\pm0.14$ &$ -0.26\pm0.08$  \\
&  1.1$<$\z$<$1.3 &        &$ 0.38\pm0.04$ &$ 1.37\pm0.17$ &$ 1.50\pm0.15$ &$ -0.13\pm0.03$ &$ 0.37\pm0.10$ &$ 1.50\pm0.14$ &$ -0.26\pm0.08$  \\
&  1.3$<$\z$<$1.5 &        &$ 0.22\pm0.05$ &$ 1.76\pm0.20$ &$ 1.83\pm0.16$ &$ -0.18\pm0.03$ &$ 0.33\pm0.12$ &$ 1.73\pm0.17$ &$ -0.34\pm0.08$  \\
\hline 

\end{tabular} 
\caption{Best 
 fitting parameters of the non linear biasing models given in eqs. (\ref{dlm})
and (\ref{tayl}). Errors do not include cosmic variance.}  

\label{table2} 
\end{center}
\end{table*}

\section{The Biasing Function up to $z\sim 1.5$}
\subsection{Results}

The numerical solutions of eq. (\ref{de}) for the $\mathcal{M}^{c}_{B}=-20+5\log h$ volume-limited 
VVDS sample are plotted in various redshift slices,
in Fig. \ref{fignlb12}  for the cases R=8 and 10 \hpcp
Note that a log-log density plot is 
used in order to emphasize the behavior 
of the biasing function in underdense regions (note that in these units 
linear biasing appears as a curved line).

The corresponding  parameters $\bl$ and  r (also computed for the whole 
flux-limited sample) are quoted in 
table \ref{table1}, together with our estimates of 
the second moment of the galaxy PDF ${\sigma_R}$ and 
of the skewness parameter S$_3$. 
Both these statistics have been computed as described in \S 5.
Also note that the values of $\sigma_8$ measured for the flux-limited sample
are consistent  with the values independently derived in \S 3.4 on the basis of
the results of the analysis of  the clustering properties of VVDS galaxies
(Paper III).

An empirical fit of the biasing function  is obtained by using  a
formula similar to the one proposed by \citet{del99}

\begin{equation} \delg(\del) = \cases{ (1+a_0) (1+\del)^{a_1} -1 & $\del
\leq 0$ \cr a_0 +a_2 \del + a_3 \del^2 & $\del>0$ }. \label{dlm} \end{equation}

\noindent which best describes the behavior of biasing in underdense regions ($\del<0$),
either the second order Taylor expansion of the density contrast of dark matter 
(Fry\& Gazta\~naga 1993) 
 
\begin{equation}
\delg=\sum_{k=0}^{2}\frac{b_k}{k!}\del^k. \label{tayl}
\end{equation}
which allows an easier comparison of our results with  other studies.
The  best fitting parameters of these non-linear approximations are quoted in table \ref{table2}.

The dependence of the shape of the biasing function on  galaxy luminosity 
is plotted in Fig. \ref{figres2}.  
Results are shown at the median depth of the VVDS 
sample (in the redshift bin 0.7$<$\z$<$0.9) where faint objects ($\mathcal{M}_{B}<-17.7+5\log h$) are still sampled. 

In Fig. \ref{biasev} we show the redshift evolution of the
linear biasing parameter $\bl$ computed over the redshift interval $0.4<z<1.5$
for both the flux and volume limited samples.
We can conclude that biasing is not changing with cosmic time for $z<0.8$, 
while there is a more pronounced evolution of biasing in the redshift interval 
[0.8,1.5]. In particular, the difference between the value of $b_L$ at redshift 
z$\sim$1.5 and z$\sim$ 0 
for a population of galaxies with luminosity $\mathcal{M}_{B}<-20+5\log h$ 
is $\Delta b_L=\sim 0.5\pm0.14$, thus 
significant  at a confidence level greater than 3$\sigma$.

In Fig. \ref{lumev} we show the dependence of the linear biasing parameter on 
galaxy luminosity. Intrinsically 
brighter galaxies are   more strongly biased than  less  
luminous ones at every redshift and the dependence of  biasing on luminosity at
\z$\sim$0.8 is  in good  agreement with what  is observed  in  the local
Universe \citep{nor01}.

Given the difference in the rest-frame colors of elliptical and
irregular galaxies and the fact that the observed I band corresponds
to bluer rest-frame bands at higher redshift, the relative fraction of
early- and late-type galaxies in our I band limited survey will change
as a function of redshift.
                                                                                
Specifically, the observed difference in the B-band luminosity
function of early- and late-types \citep{zuc}, implies that the VVDS
survey selects preferentially late-type galaxies at higher redshift.
It is known that at \z=0 late-type galaxies cluster less strongly than
early-types (\eg Giovanelli, Haynes \& Chincarini 1986, Guzzo et al.
1997, Giuricin et al. 2001, Madgwick et al. 2002, Zehavi et al. 2002),
and, thus, we might observe a variation of the amplitude of density
fluctuations at high redshifts just because the morphological
composition of our sample changes.
                                                                                
In order to disentangle the spurious morphological contribution to the
observed evolution of the global biasing function we have splitted our
sample according to rest frame colors, selecting a red
($(B-I)_0>1.5$; 849 galaxies in the 4-passes region with $z>0.7$) and a blue subsample of galaxies ($(B-I)_0<1$; 1891 galaxies with $z>0.7$).
These color cuts roughly correspond to selecting, respectively,
morphological types $\leq II$ and IV according to the classification
scheme devised by Zucca et al. 2005 for the VVDS sample.
                                                                                
Clearly, this subsample selection
does not correspond to the ideal case of a  redshift survey sampling 
galaxies according to  their rest-frame colors; however, useful information 
about differences in clustering between red and blue populations can still be inferred.

Note that the hypothesis on which the technique of comparing mass and galaxy 
density distributions is based (\S 6) can be straightforwardly generalized to 
compute the biasing between the density distributions of different galaxy types.
In particular,  we assume that 
the large scale velocities of late and early types are not dissimilar relative 
to each other \citep[as it is effectively observed at z=0 \eg][]{dek94,mar98}
\ie the two velocity fields are  noisy versions of the same 
underlying field.   

Results about the color dependence of biasing are 
summarized in table \ref{table4} and graphically presented in 
Fig. \ref{fignlbtype2}. The red sample is  systematically 
a more biased tracer of mass  than the blue one in every redshift interval 
investigated (\ie $b^r>b^b$), but the relative biasing between the two populations  
is nearly   constant ($b^{r}/b^{b}\sim 1.4\pm0.1$)

\subsection{Analysis and Discussion}

\subsubsection{Biasing for the Global Galaxy Population}

Here we examine and interpret the results derived in the previous section.
We begin by discussing the general shape of the non-linear biasing function,
for the global population, in 
different density regions. Our results can be summarized as follows:\\

i) in underdense regions  ($1+\del<$1) the local slope of the biasing function $b(\delta)$ is always 
   larger than unity even when the global slope is $b_L < 1$ (see for example Fig. \ref{figres2}).
   The fact that galaxies in low-mass density regions are always positively biased with respect to 
   the mass distribution  (\ie 
   locally $b>1$) is possibly physically caused by the fact that galaxies  do not form in very low-density mass regions, 
   \ie  below some finite mass underdensity 
\begin{figure*} 
\centering 
\includegraphics[width=18cm, angle=0]{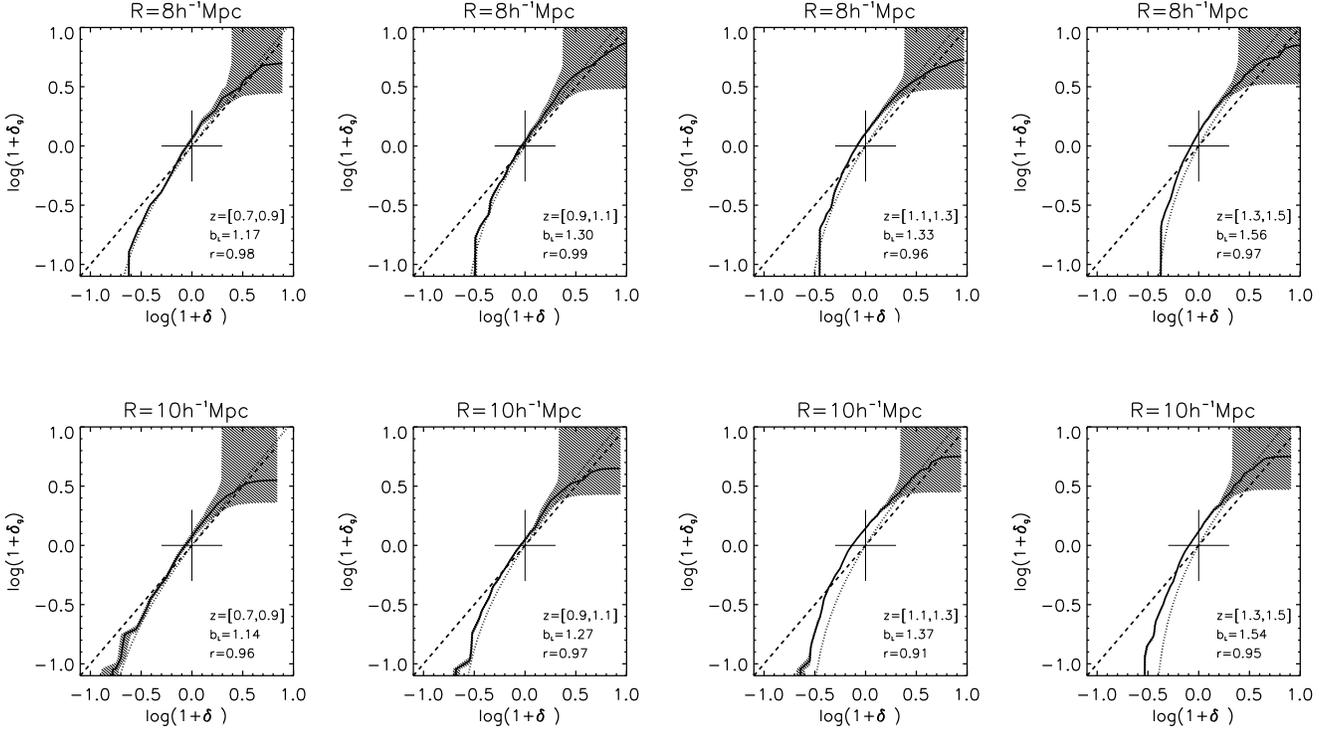}
\caption{
The observed biasing
function (solid-line) recovered for the 
density field smoothed on scales R= 8 \hpc ({\it upper panel}) and 10 \hpc ({\it lower panel})
and for different redshift bins (from left to right)
in the volume-limited VVDS sample ($\mathcal{M}^{c}_B=-20+5\log h$).
The dotted line represents the linear biasing model $\delg=\bl \del$ while 
the no-bias case ($\bl =1$) is shown with a dashed line.
The central cross is for reference and represents the $\delg=\del=0$ case. 
The shaded area represents $1\sigma$ errors  in the derived biasing 
function. Errors take into account the  noise in the observed 
galaxy PDF  ($g(\delg)$), but do not include uncertainties due to cosmic variance.}  
\label{fignlb12} 
\end{figure*}
\begin{figure*} 
\centering 
\includegraphics[width=18cm, angle=0]{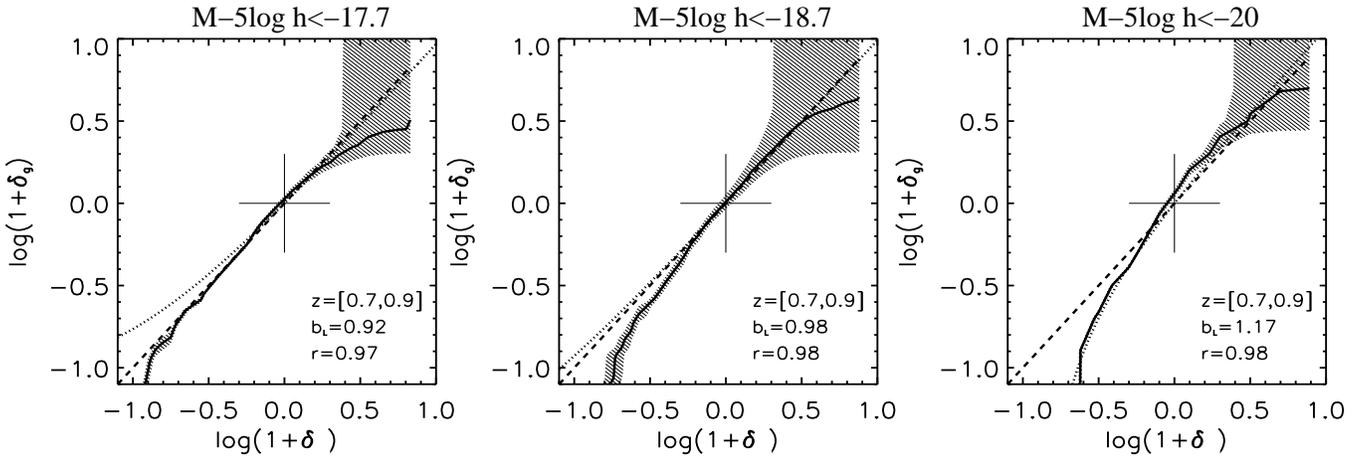}
\caption{
The biasing 
function (solid-line) on scales R= 8 \hpc and in the redshift 
interval 0.7$<$\z$<$0.9 computed for different luminosity classes.
Symbols are as in Fig. \ref{fignlb12}.}
\label{figres2} 
\end{figure*}
the galaxy formation efficiency drops to zero. 
   Using the biasing 
   relation given in eq. (\ref{dlm}) the characteristic mass density threshold $\del_c$ below which very few
   galaxies form ($\delg\leq-0.9$), can be approximated  as 

\begin{equation}
\log(1+\del_c)\sim -\frac{1+\log(1+a_0)}{a_1}
\end{equation}
   
   There is evidence  that  this mass-density threshold, characterizing regions avoided by  
   galaxies, increases as a function of redshift (see Fig. \ref{fignlb12}) and luminosity (see Fig. \ref{figres2}).
   If we consider R=8 \hpc and the redshift bin 0.7$<z<$0.9 we  see that 
   while faint galaxies seem to be present even where the mass density contrast is very low 
\begin{figure}[h] 
\includegraphics[width=9cm, angle=-0]{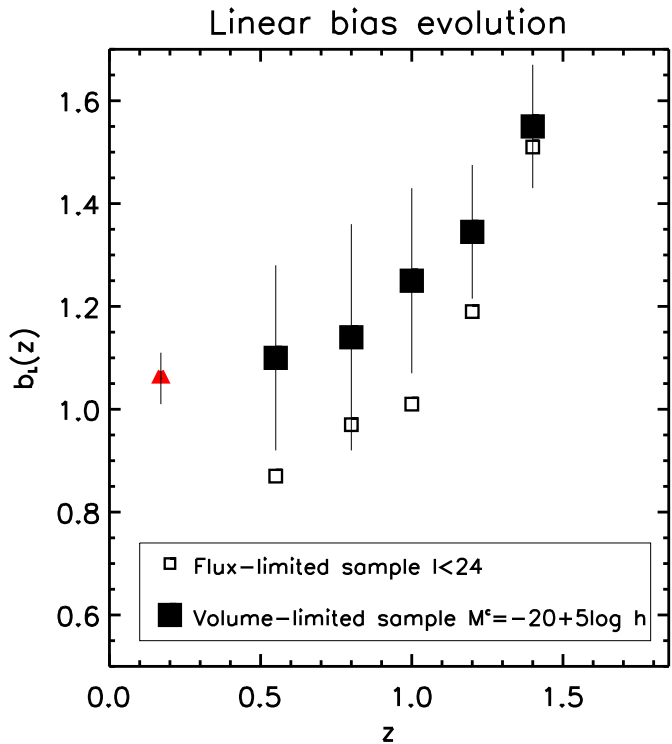} 
\caption{The redshift evolution of the 
linear biasing parameter $\bl$   for  
the volume-limited ($\mathcal{M}^{c}_{B}<$-20+5$\log$h) subsample
(filled squares)  is compared to the evolution of the  biasing parameter for the whole
flux-limited VVDS-02h sample (empty squares). Since there is no significant evidence of scale dependence
in the biasing relation, we have averaged the biasing parameters measured on 5,8, and 10 \hpc scales
in order to cover the full redshift baseline 0.4$<$\z$<$1.5.
For clarity, only the errorbars corresponding to the volume-limited sample are shown.
The triangle represents the z$\sim$0 bias 
inferred  for 2dFGRS galaxies having median $L/L^*\sim 2$ 
(\ie the median luminosity of the volume-limited VVDS sample) as explained in th text.} 
\label{biasev} 
\end{figure}
\begin{figure}[h] 
\includegraphics[width=9cm, angle=-0]{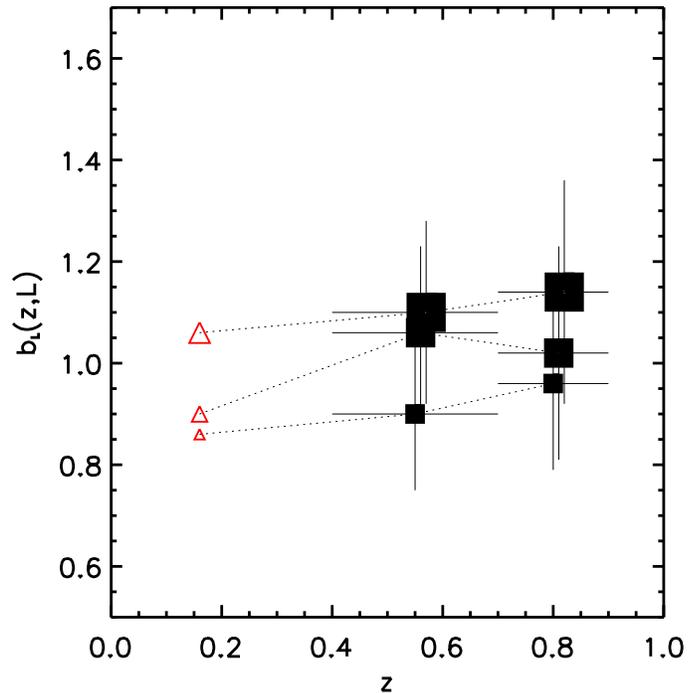} 
\caption{Comparison between the galaxy linear bias parameter 
measured in the redshift interval $0.4<z<0.9$ for 3 different luminosity classes (squares) 
and the corresponding local estimates provided by the 2dFGRS (triangles).
Points with increasing sizes correspond to three different volume-limited VVDS subsamples, i.e
$\mathcal{M}_{B}-5\log h<-17.7,<-18.7$ and $<-20$, respectively.
For clarity, squares with increasing size  have been progressively displaced  rightward to avoid crowding.
The z$\sim$0 measurements have been interpolated  
by  using the formula describing  the luminosity dependence
of the 2dFGRS  bias parameter \citep{nor01}, the bias parameter for
the 2dFGRS $L^*$ sample  (\ie $b_*=0.92$ \citep{ver02})
and the median luminosity of the three VVDS subsamples ($L/L^*$=0.52,0.82,2.0 respectively).}
\label{lumev} 
\end{figure}
(left panel of Fig. \ref{figres2}, $\log(1+\del_c)= -0.96\pm 0.09$), 
brighter galaxies do not seem to form in deep mass 
   underdensities  
   (right  panel of Fig. \ref{figres2}, $\log(1+\del_c)= -0.73\pm 0.11$). Therefore low-density regions are preferentially 
   inhabited by low luminosity galaxies. 

   Moreover the mass-density threshold below which the formation of 
   bright galaxies ($\mathcal{M}_{B}<-20+5\log h$) seems to be inhibited increases, irrespective of the scale investigated
   (see Fig. \ref{fignlb12})   as a function of redshift. On a scale R=8 \hpcv the threshold shifts from
   $\log(1+\del_c)= -0.73\pm 0.10$ at z=0.8 to  $\log(1+\del_c)\sim -0.55\pm 0.07$  at z=1.4  
   This suggests that galaxies
   of a given luminosity were tracing systematically higher mass overdensities in the early Universe, i.e,
   as time progresses, galaxy formation begins to take place also in lower density peaks.
   
ii)  Even in regions where the mass density distribution  is close to its mean value (1+$\del \sim$ 1) 
     bright galaxies are not unbiased tracer of the mass-overdensity  field (Fig. \ref{fignlb12}). 
     This can also be seen by setting $\del=0$ in eq. \ref{dlm} and noting that 
$\delg(\del=0)=a_0>0$ for both analyzed samples (flux- and volume-limited) in all redshift ranges (see table 2). 
     This result is at variance with what is expected within  the simple linear 
     biasing picture, where, by construction, $\delg(\del=0)=0$.

iii) In higher  matter-density  environments (1+$\del >1$) galaxies were progressively more biased mass tracers
     in the past, \ie the local slope $b(\del)$ systematically increases with redshift on every scale
     investigated (Fig. \ref{fignlb12}). There is some indication that, 
     at the upper tail of the mass density distribution, 
     galaxies are anti-biased with respect to mass on all scales (\ie  the local slope 
     is $b(\del)<1$ for $\del \gg 1$). Antibiasing in overdense regimes is a feature
     actually  observed in  simulations \citep[\eg][]{sbd00,som01} and expected in theoretical 
     models \citep[\eg][]{ts00}. Physically this could be due to the merging of galaxies which reduces the 
     number density of visible objects in high density regions or because galaxy formation is inhibited 
     in regions where the gas is too hot to collapse and form stars.

iv) In general the linear approximation offers a poor description of the richness of details encoded in the biasing function. As a matter of fact the linear biasing function (dotted line in Fig. \ref{fignlb12} and \ref{figres2})
    poorly describes, in many cases, the observed scaling of the biasing relation (solid line).
    At the comoving scales of R=5, 8 and 10 \hpcv non-linearities in the biasing 
    relation are typically $(1-r)\lesssim 10\%$ in the redshift range investigated.
    We find that the ratio $b_2/b_1$ between the quadratic and linear term of the 
    series approximation given in eq. (\ref{tayl}) is nearly constant in the redshift 
    range $0.7<z<1.5$ and does not depend on luminosity (i.e. it is nearly the 
    same for the flux- and volume-limited subsamples) or smoothing scale. 
    We find that, on average, $b_2/b_1\sim-0.15\pm0.04$  
    for $R=8$ \hpc and  $b_2/b_1\sim-0.19\pm0.04$ for $R=10$ \hpcp.

To facilitate comparison with other studies, which generally
focus on the linear representation
of biasing, we now discuss the properties of the linear approximation of our 
biasing function.
The general characteristics of the linear parameter $\bl$ can be summarized as follows:

v) by inspecting table \ref{table1}, we do not find any significant evidence that 
   the global value of the linear biasing parameter $\bl$  depends on the 
   smoothing scale. Any possible systematic variation, if present, is smaller than 
   the amplitude
   of our errorbars ($\sim$0.15). This scale independence in the
   biasing relation  extends into  the high redshift regimes 
   similar conclusions
   obtained in the local Universe by  the  2dFGRS  
   on scales $>$5 \hpc \citep{ver02}. 
   Moreover our results  may be interpreted as a supporting evidence for 
   theoretical arguments 
   suggesting that bias is expected to be scale-independent  on scales larger 
   than a few \hpc (\eg Mann, Peacock \& Heavens 1998, Weinberg et al. 2004).

   Since we find no evidence of scale-dependent bias, and since 
   with different R scales we
   are probing different redshift regimes, 
   in Fig. \ref{biasev} we have averaged the linear biasing parameters measured on 
   5,8, and 10 \hpc scales (values quoted in Table 1) in order to 
   follow, in a continuous way,  the redshift evolution  of the linear galaxy  biasing 
   over the larger  redshift baseline 0.4$<$\z$<$1.5. Fig. \ref{biasev} shows that 
   $\bl$ for galaxies brighter than $\mathcal{M}_B=-20+5\log h$ 
   changes from $1.10\pm0.18$ at $z \sim 0.55$ to $1.55\pm0.12$ at {\it z}$\sim 1.4$.

   An even steeper variation is observed for the biasing of the flux-limited sample, 
   indicating that biasing depends on galaxy  luminosity.
   Fig. \ref{biasev} shows that the ratio between the amplitude of galaxy fluctuations 
   and the underlying mass fluctuations declines with cosmic time. 
   This scaling is effectively predicted 
   within the framework of the peaks-biasing theoretical model \citep{kai84}.
   At early times, galaxies are expected to form at the highest peaks of 
   the density field
   since one needs a dense enough clump of baryons in order to start forming stars.
   Such high-$\sigma$ peaks are highly biased tracers of the underlying mass density field.
   According to this picture, as time progresses and the density field evolves,
   galaxy formation moves to lower-$\sigma$ peaks, nonlinear peaks become less rare
   events and thus galaxies 
   become less biased tracers of the mass density field. 
   Additional ``debiasing'' mechanisms may contribute to the observed scaling shown in Fig. \ref{biasev}. 
   It is likely that the densest regions stop forming new galaxies because their gas becomes too hot, 
   cannot cool efficiently, and thus cannot collapse and form stars \citep{bla99}. 
   As galaxy formation moves  out of the hottest (and rarest) regions of the Universe,
   the biasing decreases. Finally,
   we also note that in order to derive the biasing function we have assumed that there is no 
   difference in the velocity field of the luminous and matter components.
   After galaxies form, they are subject to the same gravitational forces as 
   the dark matter, and thus they tend to trace the dark matter distribution more closely with time as shown by \citet{dek87,fry96,teg98}.

vi) In Fig. \ref{biasev} we also show, for comparison, the value of the 2dFGRS linear biasing parameter 
   inferred at z=0.17 (the effective depth of the survey) as the ratio between  the $\sigma_8$ value measured by
   Croton et al. 2004 (in redshift-distorted space; see their Fig 3 and 4) for 
   a sample of objects with $-21<\mathcal{M}_{B}-5\log h<-20$ (which actually brackets the median luminosity of our volume limited sample $\sim 2L^*$), and the {\it rms} of mass fluctuations (in redshift-distorted space) 
   in a $\Lambda$CDM background (see \S 6). This value ($\sim1.07 \pm0.06$) is in 
   excellent agreement with what one would independently obtain by combining the linear bias 
   parameter measured by \citet{ver02} for the whole 2dFGRS ($1.04\pm0.11$) with the bias scaling 
   law recipe of \citet{nor01}, \ie $b(z=0.17,L=2L^*)=1.07\pm 0.13$.

   We can conclude that the time dependence of biasing is marginal ($db/b\sim 7\pm25\%$) for z$<$0.8
   while it is  substantial ($db/b \sim 33\pm18 \%$) in the resdhift interval [0.8-1.5]. 
    The observed time evolution   of bias is well  described by the simple
    scaling relationship  $b_L=1+(0.03\pm0.01)(1+z)^{3.3\pm0.6}$ in the interval  0$<z<$1.5.

   Assuming a linear biasing scheme, one may note that this result was already implicit in Fig. \ref{figvum}
   of \S 4. The {\it rms} fluctuations of the mass density field on a 8 \hpc scale decrease monotonically with 
   redshift by a factor of $\sim 22\%$ and $\sim 23\%$ in the redshift intervals [0.17-0.8] and [0.8-1.4], respectively; 
   thus, a nearly constant bias is  predicted in the redshift range z=[0.17-0.8] because the {\it rms} fluctuations
   of the galaxy density field are also decreasing by a factor $\sim 16\%$ in this same interval. Since, instead, 
   $\sigma_8$ of galaxies is  marginally increasing in the range z=0.8-1.4 ($d\sigma/\sigma \sim 10\%$, see table 1), 
   over this redshift baseline the biasing evolves rapidly.

vii) Bright  galaxies are more biased mass tracers than the general population 
    (see Fig. \ref{figres2}). This result confirms 
    and extends into the high redshift domain the  luminosity dependence of biasing 
    which is observed in local samples of 
    galaxies (\eg Benoist et al. 1996, Giuricin et al 2001, Norberg et al. 2001, Zehavi et al. 2002). 
    Specifically, in Fig. \ref{lumev} we show  the dependence of galaxy biasing from luminosity 
    measured  in the redshift interval 0.4$<$\z$<$0.9 using three different  volume-limited VVDS subsamples 
    (\ie $\mathcal{M}_{B}-5\log h<-17.7,<-18.7$ and $<-20$ respectively) and compare their 
  linear biasing parameters with those 
       observed locally for a sample of objects having the same  median luminosities of the VVDS subsamples
    (\ie $L/L^*$=0.52,0.82,2.0 respectively). The local estimates have been computed on the basis
    of the scaling relationship  $b/b^*$= 0.85 + 0.15 $L/L^*$ derived 
    by \citet{nor01} using the 2dFGRS sample, assuming the $b^{*}$ value given by \citet{ver02}.
    As shown above for the 
    volume-limited sample,  no significant evolution is seen up to \z $\sim$0.8 also when 
    the dependence of bias from luminosity is analyzed.

   Finally, we note that, as already discussed in \S 5,
   galaxies with the same luminosity at different redshifts may actually
   correspond to  different populations. Since, as we have shown, biasing increases 
   with luminosity also at high redshift, and since 
   the measured value of $\mathcal{M}^{*}$ for our sample at redshift z=0.4(1.5)
(Paper II) is 
   fainter(brighter) than the cut-off magnitude $\mathcal{M}^c_B=-20+5\log h$, 
   we can infer that $\bl (z)$ for a population
   of objects selected, at any given redshift,  in a narrow luminosity range around  
   $\mathcal{M}{*}(z)$
   should increase  with redshift even more than what we have 
   measured for our volume-limited sample (see Fig. \ref{biasev}).
   A more detailed  analysis of the biasing for $\mathcal{M}^*(z)$ galaxies 
   will be presented in the future, when a larger VVDS data sample 
   will be available.

\subsubsection{Biasing as a Function of Galaxy Color}

\begin{figure} 
\centering
\includegraphics[width=9.5cm, angle=0]{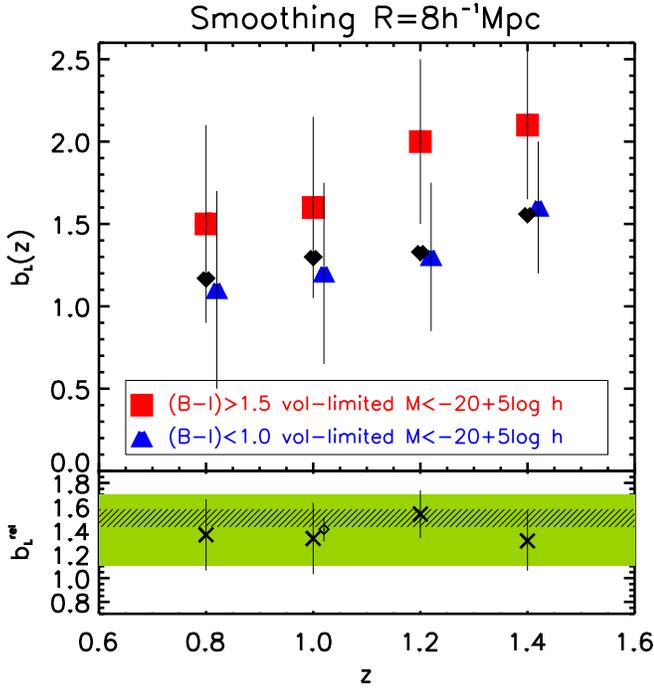} 
\caption{{\it Upper panel: }
Redshift evolution of the galaxy bias on a scale R=8 \hpc for the red (squares)
and blue (triangles) galaxies 
in the volume limited samples.  
For clarity, the triangles have been slightly displaced  rightward to avoid crowding. 
Black diamonds
represent the global bias for galaxies brighter than $\mathcal{M}^c_{B}=-20+5\log h$. 
{\it Lower panel:} the relative bias between the red and blue population ($b^{rel}=b^r/b^b$)
is shown as a function of redshift.The filled and shaded areas represents the 1$\sigma$ 
confidence region of the z$\sim$0 value for the relative bias derived  by Wilmer et al. (1998) 
and Wild et al. (2005) respectively.
The  diamond represents the relative bias measured by
Coil et al. (2004) in the redshift interval (0.7-1.35).}  
\label{fignlbtype2} 
\end{figure}

Results summarized in table \ref{table4} and 
presented in Fig. \ref{fignlbtype2} show that, 
on scales R=8 \hpcv the red sample is 
a more biased tracer of mass  than the blue one in every redshift interval.
Similarly to what we have found for the global population, 
there is some indication of a systematic increase 
as a function of redshift of the biasing of  bright red and blue objects 
even if, because of  the  
large errorbars, this trend is not statistically significant.

We can compare our results 
to the biasing measured for extremely red objects (EROS), \ie objects
with extremely red colors ($(R-K)_{vega}>5$). 
Using the results of the correlation analysis of Firth et al. (2002), we obtain, 
for their $(I-H)_{vega}>3, H_{vega}<20.5$ sample (which has a median blue luminosity 
$\mathcal{M}_B=-20.3+5\log h$),  $b_{L}^{EROS}(z\sim1.2) \sim 2.3\pm0.6$.
Considering the results of Daddi et al. (2001), who analyzed 
a sample of EROS with  $(R-K)_{vega}>5$ (which roughly corresponds to 
$(I-H)_{vega}>3$), $K_{vega}<19.2$ sample,   we conclude 
that $b_{L}^{EROS}(z\sim 1.2) \sim 4\pm1$. These values for the galaxy biasing 
are respectively $\sim 0.5$ and 1.8$\sigma$ higher than 
that measured for our sample of bright ($\mathcal{M}_B<-20+5\log h$) but
moderately red galaxies ($b_{L}^r(z \sim 1.2)=2\pm0.5$).
One may interpret this results as an indication for
the reddest objects being more strongly biased 
then moderately red galaxies of similar luminsity. 
Anyway, given the large  errorbars, the 
evidence that, at $z\sim 1.2$,
the biasing properties of these two differently selected populations are different
is not statistically significant. As a matter of fact, the values quoted above 
are also consistent with an alternative hypothesis, \ie
the strength of the EROS fluctuations with respect to the mass 
fluctuations is not  exceptional when compared to the  
density fluctuations observed in a sample of high redshift, moderately red galaxies
of similar luminosity.

\begin{table} 
\begin{center} 
\begin{tabular}{ l l l l l}
Redshift &Volume & $\bl^{rel}$&$\bl^{R}$&$\bl^{B}$   \\
range    & limited &          &         &        \\
\hline
     0.7$<$z$<$0.9 &No & $1.3\pm0.2$&$1.3\pm0.5$&$1.0\pm0.5$\\
     0.9$<$z$<$1.1 &No & $1.3\pm0.2$&$1.4\pm0.5$&$1.1\pm0.4$\\
     1.1$<$z$<$1.3 &No & $1.4\pm0.2$&$1.6\pm0.5$&$1.2\pm0.4$\\
     1.3$<$z$<$1.5 &No & $1.4\pm0.3$&$2.3\pm0.4$&$1.6\pm0.4$\\
\hline 
     0.7$<$z$<$0.9 &-20 & $1.4\pm0.3$&$1.5\pm0.6$&$1.1\pm0.6$\\
     0.9$<$z$<$1.1 &-20 & $1.3\pm0.3$&$1.6\pm0.6$&$1.2\pm0.5$\\
     1.1$<$z$<$1.3 &-20 & $1.5\pm0.2$&$2.0\pm0.5$&$1.3\pm0.4$\\
     1.3$<$z$<$1.5 &-20 & $1.3\pm0.3$&$2.1\pm0.4$&$1.6\pm0.4$\\
\hline
\hline 

\end{tabular} 
\caption{The biasing parameters for red and blue VVDS subsamples 
on a scale R=8 \hpcp}  
\label{table4} 
\end{center}
\end{table}

The specific values of the biasing parameter at each cosmic epoch are affected by 
large errors due to the sparseness of our volume-limited subsamples,  
and to the presence of cosmic variance. 
One way to bypass  uncertainties due to cosmic variance
consists in computing the relative biasing function 
$b^{rel}(\del)=b^{r}(\del)/b^{b}(\del)$
between the red and blue subsamples.
As the subsamples are drawn from the same volume, this ratio should be minimally 
affected  by the
finiteness of the volume probed by the first epoch VVDS data. 

Results about the relative biasing between galaxy of different colors 
are graphically shown in the lower panel of  Fig. \ref{fignlbtype2}, 
while  estimates of the  corresponding $\bl$ are quoted in table \ref{table4}.

We do not observe any trend in the relative biasing between red and blue volume-limited subsamples
in the redshift range  0.7$<$\z$<$1.5. Moreover, our 
best estimate $\bl^{rel}\sim 1.4\pm0.1$ is in excellent agreement with what is  found 
for nearly the same color-selected populations both
locally (Willmer et al. (1998)  found that, on a scale R=8 \hpcv $\bl^{rel}\equiv b((B-R)_{0,vega}>1.3)/b((B-R)_{0,vega}<1.3)=1.4\pm0.3$, while Wild et al. 2005 using the 2dFGRS found on the same scale $\bl^{rel}\equiv b((B-R)_{0,vega}>1.07)/b((B-R)_{0,vega}<1.07)=1.5\pm0.07$)
and at z$\sim$1 (\citet{coil} found, on a scale R=8 \hpcv  that $\bl^{rel}\equiv b((B-R)_0>0.7)/b((B-R)_0<0.7)=1.41\pm0.10$).
Thus, VVDS results  suggest that there is no-redshift dependence for  the 
relative biasing between  red and blue objects up to $z\sim 1.5$. Possible systematics
could conspire to produce the observed results;  
the linear approximation may not always captures, in an accurate way, all the 
information contained in the biasing function, and more importantly, a purely 
magnitude limited survey samples the red and blue populations at high redshift 
with a different efficiency (see discussion in \S 7.1).

In principle, the relative bias could be further studied as a function of scale.
For example, locally, there is evidence of scale dependence in the relative bias
with  the bias decreasing as scale increases (Willmer et al. 1998, Madgwick et al. 2003,
Wild et al. 2005).
However, the sample currently available is not sufficiently large  to obtain proper statistics on this effect,
although this should  be measurable from the final data set.

Finally, we note that no differences in the value of $\bl^{rel}$ are seen by 
comparing  volume-limited 
subsamples with the flux-limited one in different redshift intervals
(see table \ref{table4}). 

Thus,  we can  deduce  that  in each redshift bins
$b^{R}(\mathcal{M}_{B}<-20)/b^{R} \sim b^{B}(\mathcal{M}_{B}<-20)/b^{B}$. In other terms the biasing between  
the most luminous objects of a particular color and the global population
of objects of the same type appears to be independent of  galaxy colors
(see table \ref{table4}).

\section{Comparison with Theoretical Predictions}

In this section we compare our results
about the biasing of the $\mathcal{M}^c_B=-20+5\log h$ volume-limited, global galaxy 
sample, with predictions of different theoretical models.

Since we have found that the distribution of galaxy and mass fluctuations 
are different and the bias was systematically stronger in the past, 
we can  immediately exclude
the scenario in which galaxies trace the mass at all cosmic epochs.  
We thus consider more complex theoretical
descriptions of the biasing functions, in particular three different
pictures based on orthogonal ideas of how  evolution proceeds:
the {\it conserving}, the {\it merging}, 
and the {\it star forming} biasing models (see \eg \citet{mos98}).

In the first model the number of galaxies is conserved as a function of time
\citep{dek87,fry96}.  This model does not assume anything about the
distribution and mass of dark matter halos or their connection with
galaxies. In this scheme one assumes that galaxies are biased at birth
and then they follow the flow of matter without merging, in other
terms they behave as test particles dragged around by the surrounding
density fluctuations. 
Because the acceleration on galaxies is the same as that on the dark matter, 
the gravitational evolution after formation will tend to bring the bias 
closer to unity, as described by Fry (1996) and Tegmark \& Peebles (1998).

The evolution of the bias is given by
\citep[\eg][]{teg98} 
\begin{equation} b(z)=1+(b_{f}-1) \frac{D(z_f)}{D(z)} \label{con}
\end{equation}

\begin{figure} 
\centering
\includegraphics[width=8cm]{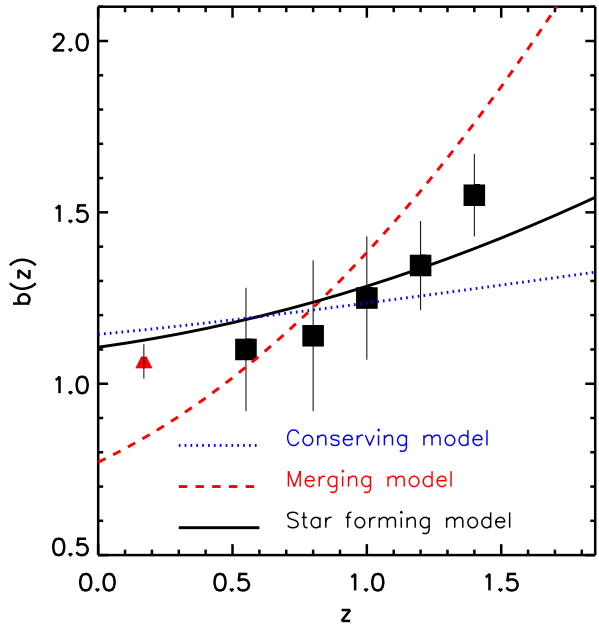} 
\caption{The redshift evolution of the linear  biasing parameter $\bl$ for the
volume-limited ($\mathcal{M}^c_{B}=-20+5\log h$) sample (see Fig 13) is compared to 
various theoretical models of biasing evolution. 
The dotted line indicates the conserving model normalized 
at $b_f(z=1.4)=1.28$, the solid and dashed lines
represent the star forming and merging models with the mass thresholds set at $3.2\cdot 10^{11}$ and $2.4\cdot 10^{12} h^{-1}$ M$_{\odot}$ respectively.}
\label{figbteo} 
\end{figure}

\noindent where $b_f$ is the bias at the formation time $z_f$.

An alternative picture for the bias evolution, which explicitly takes
into account galaxy merging, has been proposed by \cite{mow96}
who gave analytical prescriptions for computing the bias of halos
using the Press \& Schechter formalism.  

If we explicitly assume that galaxies can be
identified with dark matter halos, an 
approximate expression for the biasing  of all halos  of mass  $>M$ existing 
at redshift z (but  which collapsed at  redshift 
greater than the observation redshift, see discussion in Matarrese et al. 1997) is given
by 

\begin{equation}
b(M,z)=1+\frac{1}{\delta_c}\Big(\frac{\delta^{2}_c}{\sigma^{2}(M,z)}-1\Big)
\end{equation}

\noindent where $\delta_c\sim1.69$ is the linear overdensity of a sphere 
which collapses in an Einstein-de Sitter Universe
and $\sigma(M,z)$ is the linear {\it rms} fluctuations on scales corresponding to mass M at the
redshift of observation.

\begin{table} 
\begin{center} 
\begin{tabular}{ l l l l l}
Model &Best fitting parameters & $\chi^2/$dof   \\
\hline
Conserving   &$b_f(z=1.4)=1.28\pm0.03$& 2\\
Merging      &$M=2.4\cdot10^{12}h^{-1}$M$_{\odot}$& 5.5\\
Star forming &$M=(3.2\pm3)\cdot10^{10}h^{-1}$M$_{\odot}$ & 0.7\\
\hline
\hline 

\end{tabular} 
\caption{Best fitting parameters and the corresponding $\chi^2$ values 
for various biasing models}
\label{table5} 
\end{center}
\end{table}

The third model is also framed  within the  peaks-biasing formalism. It  assumes  that the 
distribution of galaxies with luminosity $>L$ 
is well traced by halos with mass $>$M, and predicts the biasing of objects that 
just collapsed at the redshift of observation (\eg Blanton et al. 2000). 
In this  {\it star forming} model, 

\begin{equation}
b(M,z)=1+\frac{\delta_c}{\sigma^{2}(M,z)}
\end{equation}

\noindent represents the biasing of galaxies that formed in a narrow time interval 
around redshift z (\ie galaxies which experienced recent star formation at redshift z.) 
 
Clearly the above  models, 
are based on a set of theoretical ingredients which represent a crude approximation of the
complex multiplicity of physical phenomena entering the cosmic recipe
of galaxy biasing. In this context, our goal   is to  investigate the
robustness of the simplifying assumptions on which theoretical models
are based, and explore the validity or limits of their underlying
physical motivations.

Theoretical predictions are compared to observations (VVDS data plus the 
local normalization derived from 2dFGRS data) in Fig. \ref{figbteo}.
The best fitting parameters for each model are evaluated  using a $\chi^2$ 
statistics and are quoted, together with the corresponding minimum $\chi^2$ 
value of the fit, in table 4. 

The best fitting galaxy conserving model 
is obtained when the bias at birth is $b_f(z_f=1.4)=1.28\pm0.03$ and the corresponding 
normalized $\chi^2$-value is $\chi^{2}_{N}=2$.
As shown in Fig. \ref{figbteo} the redshift evolution predicted by this model is much weaker 
than suggested by data. 
Thus, the gravitational debiasing is a physical mechanism that 
alone may not fully  explain the observed redshift evolution of the biasing, in the sense that 
it significantly underpredicts the rate of evolution. 

The redshift evolution is more pronounced in the {\it merging} model (specifically, in Fig. \ref{figbteo},
we show the bias evolution of galaxies hosted in halos having mass $M\gtrsim 2.4 \cdot 10^{12} h^{-1}M_{\odot}$). 
While this model 
successfully describes the redshift dependence of the biasing of halos \citep{mow96,som01} 
it poorly accounts for the redshift evolution of the bias of galaxies (with $\mathcal{M}_{B}\leq -20+5\log h$) 
between \z=0 and  \z=1.5 which is slower than predicted ($\chi^{2}_{N}=5.5$ for the best fitting model). 
Thus, although merging is an important mechanism for describing the evolution of
matter  clustering, our result
implies that merging processes affect galaxies in a less dramatic way than halos.
Since in the Press \& Schechter formalism halos are required to merge instantaneously in bigger units at the 
redshift of observation, our result would imply, also, that the merger time-scales 
of galaxies is different 
from that of halos. Moreover, selecting galaxies with a fixed 
luminosity threshold may not correspond, over such a wide \z range as that investigated here, to selecting 
halos above a given fixed mass threshold. In this sense our result would be suggestive of evolution
in  the mass-to-light ratio as a function of time.

In Fig \ref{figbteo} we also show 
the expected redshift evolution for the {\it star forming} model 
for halos of $M\gtrsim 3.2 \cdot 10^{10}h^{-1}$M$_{\odot}$. In this case,  
the agreement between model and observations is better ($\chi^{2}_{N}=0.7$).
Clearly this does not  mean that we are analyzing a sample of objects that 
just collapsed and formed stars at the time
they were observed; as a matter of fact the model cannot capture all the 
physical processes shaping the biasing relation. Moreover, the low value fitted 
for the mass threshold is somewhat unrealistic for the bright 
objects we are considering. Notwithstanding, Blanton et al. 2000 already noted 
that the prediction of this biasing model is not much 
different from the biasing evolution expected for the general population of
galaxies in a hydrodynamical simulation of the large scale structure. 

Our analysis  seems to suggest
the apparent need of more complex biasing models 
that better approximate the observed biasing evolution. 
Understanding our results completely, however,  
will require more discriminatory
power in the data, and, thus, a larger VVDS sample.

\section{Summary and Conclusions}

Deep surveys of the Universe provide the basic ingredients needed to
compute the  probability distribution function of galaxy
fluctuations and to constrain its evolution with cosmic time. 
The evolution of the galaxy PDF may shed light onto the general
assumption that structures grows via gravitational collapse of density 
fluctuations that are small at early time.
When this statistics is combined with analytical CDM predictions for the
PDF of mass, useful insights into the biasing function relating mass
and galaxy distributions can be obtained. 

In this paper, we have explored the
potentiality of this approach by analyzing the first-epoch data of the
VVDS survey. This is the largest, purely flux-limited sample of
spectroscopically measured galaxies, currently available  in a continuously 
connected volume and with a robust sampling  
up to redshifts z$\sim$1.5. It is worth to emphasize that the  VVDS 
is probing the high redshift domain at I$\leq$24 in the VVDS-02h-4 field 
with the same sampling  rate of pioneer surveys of the local Universe  such as the CFA
(at z$\sim$0) and, more recently, the 2dFGRS (at z$\sim$0.1).

Particular attention has been paid to assess the completeness of the
VVDS sample and  to test the statistical reliability of the PDF of VVDS
galaxy fluctuations. In particular:

{\it a)} by applying the VVDS observational selection functions to GALICS semi-analytical 
   galaxy simulations we have  explored the region of the parameter space 
   where the PDF of VVDS-like  densities traces in a
   statistically unbiased way the parent underlying PDF of the real
   distribution of galaxy overdensities.

{\it b)} we have reconstructed the VVDS galaxy density field on different
   scales R=[5,8,10] \hpc by correcting the density estimator for various VVDS
   selection functions.  The final density map for the flux-limited sample 
   has been Wiener filtered
   in order to minimize the shot-noise contribution

By studying the PDF of galaxies in the high redshift Universe we have found that 
the peak of the galaxy PDF systematically shifts to lower density contrasts as 
a function of redshift and that the 
probability of observing  underdense regions is greater at \z$\sim$0.7 than 
it was at \z$\sim$1.5. Both these effects provide strong supporting 
evidence for the standard assumption that the large scale structure is the result
of the gravitational growth of small primordial
density fluctuations in an expanding universe.

First,  within the paradigm of 
gravitational instability,
the assembling process of the large-scale structures is thought to be 
regulated by the interplay of two competing effects: the tendency of 
local self-gravity to make overdense regions collapse and the opposite
tendency of global cosmological expansion to move them apart. A key signature
of gravitational evolution of density fluctuations in an expanding Universe
is that underdense regions, experiencing the cosmological matter outflow, 
occupy a larger volume fraction  at present epoch than in the early Universe.  
Secondly,  both this effects, the peak shift and the development of a low density tail,
could indicate the existence of a time-evolving  biasing between matter 
and galaxies, since galaxy  biasing, systematically increasing with redshift, 
offers a natural mechanism 
to re-map the galaxy PDF  into progressively  higher intervals of density contrasts. 

This last interpretation is confirmed by our measurements  of the evolution properties
of the second and third moments of the galaxy PDF. We find that {\it i}) 
the {\it rms} amplitude of the fluctuations  of bright VVDS galaxies
is with good approximation constant 
over the full redshift baseline investigated. 
Specifically we have shown that, in redshift space, $\sigma_8$ for galaxies brighter 
than $\mathcal{M}_B^{c}=-20+5\log h$ has a mean value of 0.94$\pm0.07$
in the redshift interval 0.7$<$\z$<$1.5; ii)  the third moment of the PDF, 
\ie the skewness, increases with cosmic time. Its value at $z\sim1.5$ is nearly 
$2\sigma$ lower than measured locally by the 2dFGRS.
Both these results, when compared 
to predictions of linear and second order perturbation theory, unambiguously  indicate
that galaxy biasing is an  increasing function of redshift. 

Exploiting the sensitivity of the galaxy PDF 
to the specific form of the mass-galaxy
mapping,  we have derived the redshift-, density-, and
scale-dependent biasing function $b(z, \del, R)$ between galaxy and
matter fluctuations in a $\Lambda$CDM Universe, 
by analyzing the Jacobian transformation between their
respective PDFs. 
Particular attention has been paid to 
devise an optimal strategy so that the
comparison of the PDFs of mass and galaxies can be carried out in
an objective and accurate way. Specifically,
we have corrected the lognormal approximation, which describes
the mass density PDF, in order to take into account redshift
distortions induced by galaxy peculiar velocities 
at early cosmic epochs where the mapping between redshifts
and comoving positions is not linear. In this way, theoretical
predictions can be directly compared to
observational quantities derived in redshift space.

Without {\it a-priori} parameterizing the form of the biasing function, we 
have shown  its general non trivial shape, and studied its evolution
as a function of cosmic epochs. Our main results about biasing in the high redshift
Universe can be summarized as follows:

\noindent {\it i}) we detect non-linear effects in the biasing relation. 
The ratio between the quadratic and linear term of the biasing
expansion (cfr eq. \ref{tayl}) is different from zero 
at a confidence level greater than 3$\sigma$ 
in all the redshift bins and for all the smoothing scales probed.
This result confirms a  general prediction 
of CDM-based hierarchical models of galaxy formation \citep[\eg][]{sbd00,som01}. 
Such non-linear distortions of the biasing function are  not observed locally 
in the 2dFGRS sample \citep[][but see Baugh et al. 2004]{ver02}, although 
indirect evidences of a non-linear bias at $z\sim0$ exist (Benoist et al. 1999, 
Baugh et al. 2004).

\noindent {\it ii}) The biasing function
rises sharply in underdense regions (the local slope is  $b(\del)>1)$
indicating that  below some finite mass density threshold  the formation efficiency 
of galaxies brighter than $\mathcal{M}_{B}<-20+5\log h$ drops to zero. This threshold 
shifts towards higher values of the 
mass density field as the luminosity or the  redshift 
of the galaxy population increases. 

\noindent {\it iii}) We do not observe the imprints of scale-dependency in the biasing function
a behavior in agreement with  results derived from more local surveys at z$\sim$ 0 \citep{ver02}.

\noindent {\it iv}) By representing the biasing function in linear approximation,
we have found that the linear biasing parameter $b_L$ evolves with cosmic time:
it appears that we live in a special epoch in which the galaxy distribution
traces the underlying mass distribution on large scales ($b_L \sim 1$), while, 
in the past, the two fields were progressively dissimilar  
and the relative  biasing systematically higher. 
The difference between the value of $b_L$ at redshift z$\sim$1.5 and z$\sim$ 0 
for a population of galaxies with luminosity $\mathcal{M}_{B}<-20+5\log h$ 
is significant  at a confidence level greater than 3$\sigma$ 
($\Delta b_L \sim 0.5\pm0.14$). In this interval,  the essential 
characteristics of the  time evolution of the linear  
bias are well described in terms of  the phenomenological  relationship 
$b_L=1+(0.03\pm0.01)(1+z)^{3.3+0.6}$.

\noindent {\it v}) Over the redshift baseline investigated, the rate of biasing evolution is a function of redshift: 
z$\sim$0.8  is the  characteristic redshift which marks the transition
from a ``minimum-evolution'' late epoch to an early period 
where the biasing evolution for a population of $\mathcal{M}_{B}<-20+5\log h$ galaxies 
is substantial ($\sim 33\pm18\%$ between redshift 0.8 and 1.5).

\noindent {\it vi}) Brighter galaxies are   more strongly biased than  less  luminous
ones at every redshift and the dependence of  biasing on luminosity at
\z$\sim$0.8 is  in good  agreement with what  is observed  in  the local
Universe.

\noindent {\it vii}) By comparing  our results to predictions of theoretical models 
for the  biasing evolution,  we have shown that   the galaxy {\it conserving} 
model \citep{fry96} and halo {\it merging} \citep{mow96} model offer a poor 
description of our data. This result could  suggest that the gravitational 
debiasing and the hierarchical merging of halos  may not be the only physical mechanisms driving
the evolution of galaxy biasing across cosmic epochs. At variance with these results,
the {\it star forming} model (Blanton et al. 2000) seems to describe better the 
observed  redshift evolution of the linear biasing factor. 

\noindent {\it viii}) After splitting the first-epoch redshift catalog  into  red and blue volume-limited subsamples, 
we have found that  the red sample is  systematically 
a more biased tracer of mass  than the blue one in every redshift interval investigated, but 
the relative biasing between the two populations  is nearly   constant in the
redshift range 0.7$<$\z$<$1.5 ($b^{r}/b^{b}\sim 1.4\pm0.1$), and  comparable with local estimates.
Moreover, we have found that the  bright red subsample is biased with respect to
the general red population in the same way as the bright sample of blue objects 
is biased with respect to the global blue population thus indicating, 
that  biasing as a function of luminosity might be,  at first order, 
independent of colors. 

\noindent {\it ix}) Because the VVDS and various EROS samples are not yet 
large enough, the bias  of our sample
of bright and moderately red objects at z$\sim$1 is not statistically dissimilar from that expected for EROS of similar 
luminosity, even if the EROS biasing appears to be systematically larger. 

One key aspect of this paper is the measure of evolution in the distribution
properties of galaxy overdensities from a continuous volume sampled with the 
same selection function over a wide redshift baseline.
As our volume sampled is still limited, errors on the analysis presented in this paper
are dominated by cosmic variance. 
The technique presented here will be applied to a larger sample as the VVDS 
observational program  progresses.

\section*{Acknowledgments}

We would like to acknowledge useful discussions with
A. Dekel, R. Giovanelli, and L. Moscardini. We also thank S. Andreon and 
J. Afonso for their  useful comments on the paper.
This research has been developed within the framework of the VVDS consortium
and it has been partially supported by the CNRS-INSU and its Programme
National de Cosmologie (France), and by the Italian Ministry (MIUR) grants
COFIN2000 (MM02037133) and COFIN2003 (num.2003020150).
CM also acknowledges financial support from the Region PACA.
The VLT-VIMOS observations have been carried out on guaranteed
time (GTO) allocated by the European Southern Observatory (ESO) to the
VIRMOS consortium, under a contractual agreement between 
the Centre National de la Recherche Scientifique of France, heading 
a consortium of French and Italian institutes, and ESO, to design, manufacture
and test the VIMOS instrument. 
We thanks the GALICS group for privileged access to their semi-analytic
simulations. The mass simulations used 
in this paper were carried out by the Virgo Supercomputing Consortium
using computers based at the Computing Centre of the Max-Planck
Society in Garching and at the Edinburgh parallel Computing Centre.

\bigskip \bigskip \section*{Appendix A}

Here we describe the application of the Wiener filtering technique to 
deconvolve the noise signature  from the VVDS density map.
We de-noise data in Fourier space, noting, however, that an equivalent
filtering can be directly applied in real space \citep[\eg][]{ryb,
zar}.

 Let us assume that the observed smoothed density field
$\delo(\blx)$, and the true underlying density field $\delt(\blx)$,
smoothed on the same scale, are related via

\begin{equation} \delo(\blx) = \delt(\blx) + \epsilon(\blx),
\end{equation}

\noindent where $\epsilon(\blx)$ is the local contribution from shot
noise (see eq. (\ref{shot})).  The Wiener filtered density field, in
Fourier space, is

\begin{equation} \tdelf(\blk) = \mathcal{F}(\blk) \tdelo(\blk),
\end{equation}

\begin{equation} \mathcal{F}(\blk) = { \vev{\tdelt^2(\blk)} \over
\vev{\tdelt^2(\blk)} + (2\pi)^{3} P_{\epsilon}(\blk)}\ .
\label{Wiener} \end{equation}

where brackets denote statistical averages and where
$P_{\epsilon}(\blk)=(2\pi)^{-3} \vev{|\teps^2(\blk)|}$ is the power
spectrum of the noise. Assuming ergodic conditions for the
noise, we can  derive its power spectrum as
$P_{\epsilon}(\blk)=(2\pi)^{-3} |\teps(\blk)|^2.$

The Power spectrum of the underlying theoretical density distribution
of galaxies, smoothed with the window F and taking into account the
VVDS geometrical constraints, can be derived from 
eq. (\ref{defdg}). Specifically 
he theoretical overdensity field smoothed on a certain scale $R$,
which is sampled by an idealized survey with no selection functions,
is 

\begin{equation} \displaystyle \del_{g}(\blr,R) \equiv
\frac{1}{\overline{\rho}}\sum_p
\del^{D}(\blr-\blr_p)*F\Big(\frac{|{\bf r-r_p}|}{R}\Big)- 1
\label{defdg2} \end{equation}

\noindent If we assume that this density field is periodic on same
volume V, having, for example, the same geometry of the VVDS survey,
its Fourier transform is

\begin{equation}
\tdelt(\blk)=\frac{1}{\overline{\rho}}\sum_i\frac{n_i}{V} e^{i{\bf k
\cdot r_i}}\fk-\frac{1}{V}\int e^{i{\bf k \cdot r}}d^3\blr \label{fad}
\end{equation}

\noindent where the mean theoretical galaxy density $\overline{\rho}$
may be estimated averaging over sufficiently large volumes the VVDS
data corrected for selection functions and sampling rate (see
eq. (\ref{defrg})).

In eq. (\ref{fad}), $n_i$ represent the occupation numbers of the
infinitesimal cells $d\blr_i$ in which the VVDS volume can be
partitioned ($n_i$=0 or 1) and the sum is intended over all the cells
of the survey volume.  The Fourier transform of the smoothing window
function F with which the discontinuous galaxy density field is
regularized is, in the case of a Top-Hat spherical smoothing filter,

\begin{equation} 
\fk=3 \Big[ \frac{sin (kR)}{(kR)^3}-\frac{cos(kR)}{(kR)^2}\Big].
\label{ftth}
\end{equation}

In order to obtain an estimate of the quantity $\vev{\tdelt^2(\blk)}$
which enters the Wiener filter definition (eq. \ref{Wiener}), we
compute the ensemble average of the squares of the modulus of the
Fourier amplitudes (eq. \ref{fad}) and obtain

\begin{eqnarray*}
\vev{\tdelt(\blk)\tdelt^*(\blkp)}&=\frac{1}{(\overline{\rho}
V)^2}\Big[\sum_i \sum_j \vev{n_i n_j} e^{i (\blk \cdot \blr_i-\blkp \cdot
\blr_j)} \fk \fskp+ \\ & +\Wk \Wskp -\overline{\rho} V \sum_i
\vev{n_i} e^{i \blk \cdot \blr_i} \fk \Wskp+\\ & - \overline{\rho}V \sum_j
\vev{n_j} e^{i \blkp \cdot \blr_j} \fskp \Wk\Big] \end{eqnarray*}

\noindent where 

\begin{equation} W(\blk)=\frac{1}{V}\int e^{i \blk
\cdot \blr}d^3\blr \label{swf} \end{equation}

The mean value of the occupation number is given by (Peebles 1980)

\[ \vev{n_i n_j}=\overline{\rho}^2 d^3\blr_i d^3\blr_j [1+\xi(
\blr_i-\blr_j)] \]

\[ \vev{n_i}^2=\vev{n_i}=\overline{\rho} \, d^3\blr_i \]

\noindent where the correlation function may be expressed via the
Fourier conjugates

\begin{equation} \xi(\blr_{ij})=\frac{1}{(2\pi)^3}\int d^3\blk
P(\blk)e^{-i {\blk \cdot \blr_{ij}}} \end{equation}

\begin{equation} P(\blk)=\int d^3\blr \xi(\blr)e^{i \blk \cdot r}
\label{pow} \end{equation}

We convert the sums into integrals over the occupation cells taking
into account the specific VVDS geometry and obtain

\begin{eqnarray} \vev{\tdelt^2(\blk)}&=|\fk|^2\int d^3\blkp
P(\blkp)G(\blk-\blkp)+\\ & +|\fk^2|(\overline{\rho} V)^{-1} +
|\Wk|^2(1-\fk)^2 \label{res} \end{eqnarray}

\noindent were

\begin{equation} G(\blk-\blkp)=\frac{1}{(2\pi)^3}|W(\blk-\blkp)|^2
\end{equation}

Note that, in the idealized conditions of a survey of infinite spatial extension
($V \rightarrow \infty$) and no smoothing applied to data the previous
expression reduces to

\begin{equation} \vev{\tdelt^2(\blk)}=(2\pi)^3
P(\blkp)\del^D(\blk-\blkp). \end{equation}

Given the quasi pencil-beam nature of the first season VVDS data, however,
the sample of the $\del$ field can be described, with good
approximation, as confined to a cylinder of length L (aligned along
the redshift direction) and radius R. In this case the Fourier transform 
of the survey window function (cfr. eq. \ref{swf}) is given by

\begin{equation} \Wk=j_0\Big( \frac{L}{2}k_{\parallel}\Big) \,2
\frac{J_1 (k_{\perp}R)}{k_{\perp}R} \end{equation}

\noindent where $j_{0}$ and $J_1$ are the spherical and first kind
Bessel functions and where $k=\sqrt{k_{\parallel}^2+k_{\perp}^2}$.

The convolution integral on the right hand side of equation \ref{res} can then be
evaluated as follows  (\eg Kaiser \& Peacock 1991, Fisher et
al. 1993)

\[\int d^3\blkp P(\blkp)G(\blk-\blkp)= \frac{2}{\pi
V}\int_{-\infty}^{\infty} dy j_{0}^2\Big(y-k\frac{L}{2}\Big)F(y) \]

\noindent where V=$\pi R^2L$ is the volume of the cylinder and where

\begin{equation} F(x)=\int_{0}^{\infty}
P\Bigg(\frac{\sqrt{x^2+x'^2}}{R}\Bigg)\frac{J_{1}^{2}(x')}{x'}dx'.
\end{equation}

Note that the theoretically expected variance of the density field  in the VVDS volume 
is derived in real comoving space. Thus, we have corrected  the theoretical predictions 
(cfr. eq. \ref{res}) in order  to take into account redshift space distortions 
induced by galaxy peculiar  velocities (see discussion in \S 5).

\begin{figure} \centering \includegraphics[width=9.2cm, angle=0]{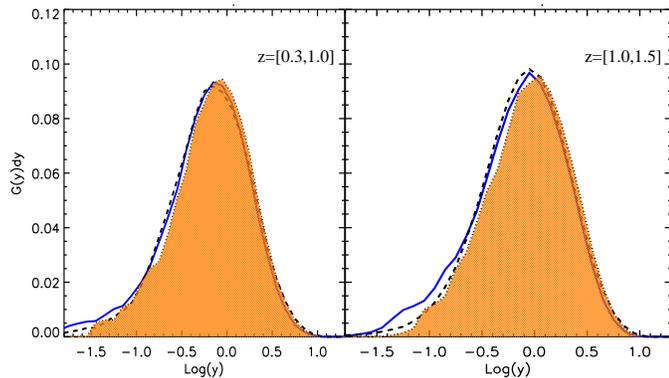}
\caption{The overdensity distribution reconstructed on a scale R=8 \hpc 
and in two different redshift intervals
using the GALICS semi-analytical simulation (see \S 4.1) is plotted 
as a function of $\log (1+\delg)$. 
The dashed line represent  the distribution of overdensities
traced by simulated  galaxies with $\mathcal{M}<-15+5\log h$. 
The shaded area (dotted line) 
represent the {\it observed} overdensity distribution \ie the distribution 
recovered after applying to the simulation 
the VVDS flux limit (I$\leq$24) and all the underlying 
VVDS instrumental selection effects (see \S 4.1).
The solid line represent the distribution of density contrasts after correcting 
the {\it observed} distribution with the Wiener technique.}
\label{wienfig} \end{figure}

We compute the galaxy density field on a regular Cartesian grid of
spacing 0.5 \hpc using the smoothing scheme presented in eq. (\ref{defdg}).
The resulting 3D density map is then Fourier transformed in redshift slices having 
line-of-sight dimensions d\z=0.1. This partition strategy is implemented in order 
to describe consistently, using cylindrical approximations, the deep survey volume 
of the VVDS (whose  comoving transversal
dimensions are an increasing function of distance.)
The Wiener filter at
each wave-vector position is then computed by using eq. (\ref{res}). Finally, 
as described in \S 3, we select only 
the Wiener filtered density fluctuations  recovered in spheres having at least 70$\%$ of their volume 
in the 4-passes, VVDS-02h-4 field.

In Fig. \ref{wienfig} we use the GALICS  semi-analytical simulation (see \S 4.1), to
show the effect of the Wiener filter on the   
reconstructed galaxy density field. As Fig. \ref{wienfig} shows, the net effect 
of the correction is to shift towards low-values the  density contrasts
having low signal-to-noise ratio. By definition, in fact, 
F is always smaller than unity. Fig. \ref{wienfig} 
shows that i) at the same density, the effects of the correction are bigger 
at high redshift where the density field is noisier due to 
the increasing sparseness of a flux-limited sample, and ii) at the same redshift 
the Wiener filter mostly affects the low-density tail of the distribution 
where the counts within the TH window are small.
It is also evident from Fig. \ref{wienfig} that, in the density interval 
$-1<\log (1+\delg)<1$,
the Wiener filtered distribution
offers a better approximation of the underlying PDF, than 
the observed (uncorrected) overdensity distribution.

\label{lastpage}

\end{document}